# The Foundation of a Generic Theorem Prover

Lawrence C Paulson
Computer Laboratory
University of Cambridge

**Abstract**

Isabelle [28, 30] is an interactive theorem prover that supports a variety of logics. It represents rules as propositions (not as functions) and builds proofs by combining rules. These operations constitute a meta-logic (or 'logical framework') in which the object-logics are formalized. Isabelle is now based on higher-order logic — a precise and well-understood foundation.

Examples illustrate use of this meta-logic to formalize logics and proofs. Axioms for first-order logic are shown sound and complete. Backwards proof is formalized by meta-reasoning about object-level entailment.

Higher-order logic has several practical advantages over other meta-logics. Many proof techniques are known, such as Huet's higher-order unification procedure.

**Key words:** higher-order logic, higher-order unification, Isabelle, LCF, logical frameworks, meta-reasoning, natural deduction

# Contents

# 1 History and overview

The dominance of classical first-order logic is challenged every year by something new. Scott's Logic of Computable Functions appeared in 1969 and attracted the interest of Robin Milner. Milner built a proof checker but found it impossibly tedious for proofs of any length. He later developed Edinburgh LCF, a proof checker that was programmable [26].

Edinburgh LCF's meta-language (ML) does not merely execute obvious command sequences. ML gives a general representation of logic. Terms and formulae are computable data, as are theorems. Each inference rule is a function from theorems to theorems. A theorem can be built only by applying rules to existing theorems.

While forwards proof is fundamental, it is often preferable to work backwards from a goal. Each inference rule maps premises to conclusion; its 'inverse' is a function (called a *tactic*) mapping a goal to subgoals. A tactic also returns a *validation*: a function from theorems to a theorem. When the backwards proof is finished, applying the validation functions performs the forwards proof, and yields the desired theorem. *Tacticals* operate on tactics, expressing control structures such as sequential or repetitive application of a tactic.

ML's secure type checking ensures soundness: theorems are constructed only by rules. Exceptions signal when a rule or tactic is wrongly applied. Still, a tactic can be *invalid*: promising more than it can deliver. If its validation function is wrong then the final forwards proof will not yield the theorem that was expected. Tactics should be built using tactics and tacticals known to be valid. The LCF environment grows with use: rules can be composed as functions; tactics can be combined by tacticals.

By 1986, Edinburgh LCF's techniques had spread to several systems [29], Standard ML had become a language in its own right, and Isabelle reached a usable form. Isabelle was intended to allow LCF-style proofs in various logics, using a representation of logic that did not require writing a function for each rule. In Isabelle-86, a rule was represented by a Horn clause; rules could be combined to build proofs [28]. (See also de Groote [14].) Since forwards and backwards proof were simply styles of proof construction, there was little difference between rules and tactics. Each state of a backwards proof was a derived rule:

$$\frac{\textit{subgoal} \quad \cdots \quad \textit{subgoal}}{\textit{original goal}}$$

Isabelle's tactics and tacticals could be used like LCF ones though they were based on totally different principles.

The initial concern was with implementation problems. Quantifiers required a syntax involving the typed $\lambda$-calculus, while the joining of rules required unification; these together required higher-order unification [19]. I experimented with many different ways of enforcing the variable conditions of quantifier rules. Eventually Isabelle-86 supported many logics: Martin-Löf's Type Theory, intuitionistic and classical sequent calculi, Zermelo-Fraenkel set theory. It was implemented in Standard ML.

Isabelle-86 was based on a naïve calculus of rules. It left many questions open, such as whether the following rules should be regarded as distinct:

$$\frac{A \quad B}{C} \qquad\qquad \frac{B \quad A}{C} \qquad\qquad \frac{A \quad B \quad A}{C}$$

The treatment of quantifiers seemed particularly unclear.

Many people have developed calculi for mathematical reasoning [7, 9, 15, 25]. A calculus of logics is often called a *logical framework*; I prefer to speak of a *meta-logic* and its *object-logics*. Isabelle-86 required a precise meta-logic suited to its aims and methods. A fragment of higher-order logic (called $\mathcal{M}$ here for 'meta') now serves this purpose. Implication expresses entailment; universal quantification expresses schematic rules and general premises; equality expresses definitions. The meta-logic extends that of Isabelle-86: it can express things like 'adding the double negation rule to intuitionistic logic entails the excluded middle.'

Isabelle constructs proofs through deductions in $\mathcal{M}$. The paper presents these methods formally in order to clarify theoretical issues such as soundness. Here is an outline of the paper:

Section 2 presents the meta-logic $\mathcal{M}$: intuitionistic higher-order logic.

Section 3 formalizes the natural deduction rules for intuitionistic propositional logic in $\mathcal{M}$. The formalization is shown sound and complete by induction on normalized proofs.

Section 4 describes how meta-level reasoning expresses object-level backwards proof, with examples from propositional logic.

Section 5 considers object-logics with quantifiers. The natural deduction system is extended to intuitionistic first-order logic and again shown sound and complete.

Section 6 extends the backwards proof methods to handle quantifiers and unification, with examples from first-order logic. The representation of eigenvariables differs from Isabelle-86.

Using Hilbert's $\epsilon$-operator, Section 7 recovers the Isabelle-86 representation and compares it with its alternative.

Section 8 describes how an implementation (the latest is called Isabelle-88) was obtained from Isabelle-86.

Section 9 concludes with a discussion of related work.

Isabelle's treatment of backwards proof has unique advantages over LCF's. An Isabelle proof state is formalized by a meta-theorem; there are no validations. The subgoals are guaranteed sufficient to obtain the final goal. Unification is naturally accommodated. The techniques are presented within $\mathcal{M}$, but resemble those of Isabelle-86 and may apply to other meta-logics.

Most of the examples below have been tested using Isabelle. They cover only propositional and first-order logic, but illustrate general techniques. Besides, the point is not to handle every esoteric logic. It is that mathematics requires a living language. Definitions extend its syntax; theorems extend its primitive modes of reasoning. In Zermelo-Fraenkel set theory the Cartesian product is defined by the power set, union, pairing, and separation axioms. 'Obvious' properties like

$$\frac{a \in A \qquad b \in B}{\langle a, b\rangle \in A \times B}$$

require tedious proofs. A mathematician, having performed the proofs, would treat Cartesian product like a primitive with the obvious properties as new rules of inference. Isabelle is an attempt to support this style.

# 2 The meta-logic $\mathcal{M}$

Alonzo Church developed higher-order logic, also called HOL or simple type theory. It is based on the typed $\lambda$-calculus [16]. Gordon [13] built his HOL theorem prover from LCF; another theorem prover is TPS [2].

Andrews [1] has written a book covering higher-order logic. Here is a brief sketch of a fragment called $\mathcal{M}$, which will be our meta-logic.

## 2.1 Syntax of the meta-logic

The *types*[1] consist of basic types and function types of the form $\sigma \to \tau$. Let the Greek letters $\sigma$, $\tau$, and $\upsilon$ stand for types.

The *terms* are those of the typed $\lambda$-calculus — constants, variables, abstractions, combinations — with the usual type constraints. Let $a$, $b$, and $c$ stand for terms, using $f$, $g$, and $h$ for terms of function type. Typical bound variables will be $x$, $y$, and $z$. Write $a : \sigma$ to mean '$a$ has type $\sigma$.'

The basic types and constants depend on the logic being represented. But they always include the type of propositions, *prop*, and the logical constants of $\mathcal{M}$. A *formula* is a term of type *prop*. Let $\phi$, $\psi$, and $\theta$ stand for formulae. The implication $\phi \Rightarrow \psi$ means '$\phi$ implies $\psi$.' The universally quantified formula $\bigwedge x.\phi$ means 'for all $x$, $\phi$ is true,' where $x$ ranges over some type $\sigma$. The equality $a \equiv b$ means '$a$ equals $b$.'

The symbols $\Rightarrow$, $\bigwedge$, and $\equiv$ have been chosen to differ from symbols of *object-logics*: those to be represented in $\mathcal{M}$. In an object-logic presented below the corresponding symbols are $\supset$, $\forall$, and $=$. The words 'meta-implication,' 'meta-equality,' 'meta-formula,' 'meta-theorem,' 'meta-rule,' etc., refer to expressions of $\mathcal{M}$.

Quantification involves $\lambda$-abstraction. For every type $\sigma$, there is a constant $\bigwedge_\sigma$ of type $(\sigma \to prop) \to prop$. The formula $\bigwedge x.\phi$, where $x$ has type $\sigma$, abbreviates $\bigwedge_\sigma(\lambda x.\phi)$. Using $\lambda$-conversions every quantification can be put into the form $\bigwedge_\sigma(f)$, more readably $\bigwedge x.f(x)$, where $f$ is a term of type $\sigma \to prop$. Abstraction also expresses quantifiers in object-logics, as we shall see in Section 5.

[1]Sometimes called *arities*, following Martin-Löf, to avoid confusion with ML types or object-level types.

## 2.2 Syntactic conventions

The application of $a$ to the successive arguments $b_1, \ldots, b_m$ is written $a(b_1, \ldots, b_m)$:

$$a(b_1, \ldots, b_m) \quad \text{abbreviates} \quad (\cdots (ab_1) \cdots b_m)$$

In the absence of parentheses, implication ($\Rightarrow$) groups to the right. Let $\Phi$, $\Psi$, and $\Theta$ stand for lists of formulae. Implication can also be written for such lists: if $\Phi$ is the list $[\phi_1, \ldots, \phi_m]$, then

$$\left.\begin{array}{r} \phi_1 \Rightarrow \cdots \Rightarrow \phi_m \Rightarrow \psi \\ [\phi_1, \ldots, \phi_m] \Rightarrow \psi \\ \Phi \Rightarrow \psi \end{array}\right\} \text{ each abbreviate } \quad \phi_1 \Rightarrow (\cdots \Rightarrow (\phi_m \Rightarrow \psi) \cdots)$$

One $\lambda$ or quantifier does the work of many:

$$\left.\begin{array}{r} \lambda x_1 \ldots x_m \,.\, a \\ \bigwedge x_1 \ldots x_m \,.\, \phi \end{array}\right\} \text{ abbreviates } \left\{\begin{array}{l} \lambda x_1 \,.\, \ldots \lambda x_m . a \\ \bigwedge x_1 \,.\, \ldots \bigwedge x_m . \phi \end{array}\right.$$

The scope of a $\lambda$ or quantifier extends far to the right:

$$\left.\begin{array}{r} \lambda x \,.\, f(x, g(x)) \\ \bigwedge x \,.\, \phi \Rightarrow b \equiv c \end{array}\right\} \text{ abbreviates } \left\{\begin{array}{l} \lambda x \,.\, (f(x, g(x))) \\ \bigwedge x \,.\, (\phi \Rightarrow (b \equiv c)) \end{array}\right.$$

A *substitution* has the form $[a_1/x_1, \ldots, a_k/x_k]$, where $x_1, \ldots, x_k$ are distinct variables and $a_1, \ldots, a_k$ are terms. If $b$ is an expression and $s$ is the substitution above then $bs$ is the expression that results from simultaneously replacing every free occurrence of $x_i$ by $a_i$ in $b$, for $i = 1, \ldots, k$ (of course $a_i$ must have the same type as $x_i$). Substitution must be carefully defined to avoid capture of free variables.

Substitutions are not part of $\mathcal{M}$ itself. The term $f(a)$ indicates function application, not substitution. The $\beta$-reduction law, namely $((\lambda x.b)(a)) \equiv b[a/x]$, expresses substitution at the object-level.

## 2.3 Semantics of the meta-logic

Higher-order logic is a language for writing formal mathematics. It can be justified on intuitive grounds, or else we can demonstrate its consistency by constructing a standard model in set theory.

Every type denotes a non-empty set. Given sets for each basic type, the interpretation of $\sigma \to \tau$ is the set of functions from $\sigma$ to $\tau$. A closed term of type $\sigma$ denotes a value of the corresponding set. Given a value for each constant, $\lambda$-abstractions denote functions.

The type *prop* denotes a set of truth values. Classical logic uses $\{\mathrm{T}, \mathrm{F}\}$; topos theory provides various intuitionistic interpretations [22]. The logical constants ($\bigwedge_\sigma$, $\Rightarrow$, and $\equiv_\sigma$) denote appropriate truth-valued functions.

**Remark.** Requiring all types to be non-empty permits this simple inference system, where

$$(\bigwedge x \,.\, \bigwedge \theta \,.\, \theta) \Rightarrow (\bigwedge \theta \,.\, \theta)$$

is a theorem. (Here $\theta$ is a bound variable of type *prop*!) If the type of $x$ were empty then $\bigwedge \theta \,.\, \theta$ would be true; every formula would be true; the logic would be inconsistent.

Lambek and Scott [22, pages 128–132] present an inference system for higher-order logic allowing empty types.

## 2.4 Inference rules

The constant symbols include, for every type $\sigma$,

$$\begin{aligned} \Rightarrow &\;:\; prop \to (prop \to prop) \\ \bigwedge_\sigma &\;:\; (\sigma \to prop) \to prop \\ \equiv_\sigma &\;:\; \sigma \to (\sigma \to prop) \end{aligned}$$

The *implication* rules are $\Rightarrow$-introduction and $\Rightarrow$-elimination:

$$\frac{\begin{matrix}[\phi] \\ \psi\end{matrix}}{\phi \Rightarrow \psi} \qquad\qquad \frac{\phi \Rightarrow \psi \qquad \phi}{\psi}$$

These are natural deduction rules; $\Rightarrow$-introduction discharges the assumption $\phi$. In most other rules, the conclusion depends on all assumptions of the premises.

The *universal quantification* rules are $\bigwedge$-introduction and $\bigwedge$-elimination:

$$\frac{\phi}{\bigwedge x.\phi} \qquad\qquad \frac{\bigwedge x.\phi}{\phi[b/x]}$$

These are also called *generalization* and *specialization*. The generalization rule is subject to the eigenvariable condition that $x$ is not free in the assumptions.

The *equality* rules are reflexivity, symmetry, and transitivity:

$$a \equiv a \qquad\qquad \frac{a \equiv b}{b \equiv a} \qquad\qquad \frac{a \equiv b \qquad b \equiv c}{a \equiv c}$$

The $\lambda$-conversions are $\alpha$-conversion (bound variable renaming), $\beta$-conversion, and extensionality:

$$(\lambda x.a) \equiv (\lambda y.a[y/x]) \qquad\qquad ((\lambda x.a)(b)) \equiv a[b/x] \qquad\qquad \frac{f(x) \equiv g(x)}{f \equiv g}$$

The $\alpha$-conversion axiom holds provided $y$ is not free in $a$. Extensionality holds provided $x$ is not free in the assumptions, $f$, or $g$. Extensionality is equivalent to $\eta$-conversion, namely $(\lambda x.f(x)) \equiv f$ where $x$ is not free in $f$ (see Hindley and Seldin [16, pages 72–74]).

The *abstraction* and *combination* rules are

$$\frac{a \equiv b}{(\lambda x.a) \equiv (\lambda x.b)} \qquad \frac{f \equiv g \quad a \equiv b}{f(a) \equiv g(b)}$$

Abstraction holds provided $x$ is not free in the assumptions.

Logical equivalence means equality of truth values:

$$\frac{\begin{array}{cc}[\phi] & [\psi] \\ \psi & \phi\end{array}}{\phi \equiv \psi} \qquad \frac{\phi \equiv \psi \quad \phi}{\psi}$$

The typed $\lambda$-calculus satisfies the strong normalization and Church-Rosser properties [16]. Thus repeatedly applying $\beta$ and $\eta$-reductions always terminates. The reductions can take place in any order; the resulting normal form will be the same up to $\alpha$-conversion. To summarize:

**Theorem 1** *Every term can be reduced to a normal form that is unique up to $\alpha$-conversion.*

**Remark.** Because of normal forms, equality is decidable in the typed $\lambda$-calculus — but not in higher-order logic. The normal form does not take account of the logical rules. No effective procedure can reduce every theorem to some unique true formula.

There is also a normalization procedure for HOL proofs. This plays a crucial role in demonstrating that an object-logic is faithfully expressed.

# 3 Representing intuitionistic propositional logic

To represent an object-logic in Isabelle we extend the meta-logic with types, constants, and axioms. A simple example is intuitionistic propositional logic (IPL).

To represent the syntax of IPL, introduce the basic type *form* for denotations of formulae. Introduce the constant symbols

$$\begin{array}{rcl} \bot & : & \mathit{form} \\ \&, \lor, \supset & : & \mathit{form} \to (\mathit{form} \to \mathit{form}) \\ \mathrm{true} & : & \mathit{form} \to \mathit{prop} \end{array}$$

Variables of type *form* include $A$, $B$, and $C$.

Object-sentences are enclosed in double brackets $[\![\,]\!]$. The meta-formula $[\![A]\!]$ abbreviates $\mathrm{true}(A)$ and means that $A$ is true. Keeping the types *form* and *prop* distinct avoids presuming that truth-values of the object-logic are identical to those of the meta-logic. To avoid confusing these logics, let us use distinctive terminology. There is a meta-rule called $\Rightarrow$-elimination. The similar object-rule is called the $\supset$E rule, while the corresponding meta-axiom is called the $\supset$E axiom.

The natural deduction rules (Figure 1) of intuitionistic logic are represented by meta-level axioms (Figure 2). The resulting extension of $\mathcal{M}$ is called $\mathcal{M}_{\text{IPL}}$. The outer quantifiers of meta-axioms will often be dropped.

The new symbols have the usual interpretations. Let the type *form* denote a set of truth values such that &, $\vee$, $\supset$, and $\bot$ have their intuitionistic meanings [11, Chapter 5]. The axioms are true under this semantics: for example, if $A$ is true and $B$ is true then $A \,\&\, B$ is true. Meta-implication ($\Rightarrow$) expresses the discharge of assumptions. The $\supset$I axiom says that if the truth of $A$ implies the truth of $B$, then the formula $A \supset B$ is true.

The resemblance between the meta-level axioms and the rules should be regarded as a happy coincidence. An axiom formalizes not the syntax of a rule but its semantic justification. The resemblance diminishes in first-order logic (Section 5). The formalization of modal logic by Avron et al. [3] (in their meta-logic) reflects Kripke semantics rather than the syntax of the rules.

An obvious question is whether the object-logic is faithfully represented. The definition below is oriented towards natural deduction: it concerns entailments rather than theorems.

**Definition 1** Let $L$ be a logic and $A_1$, …, $A_m$, $B$ be formulae of $L$. Let $\mathcal{M}_L$ be a meta-logic obtained from $\mathcal{M}$ by adding types, constants, and axioms. Suppose that $[\![-]\!]$ is a function mapping each formula $A$ of $L$ to a meta-formula $[\![A]\!]$ of $\mathcal{M}_L$. Then say

- $\mathcal{M}_L$ is *sound for* $L$ if, for every $\mathcal{M}_L$-proof of $[\![B]\!]$ from $[\![A_1]\!], \ldots, [\![A_m]\!]$, there is an $L$-proof of $B$ from $A_1, \ldots, A_m$.
- $\mathcal{M}_L$ is *complete for* $L$ if, for every $L$-proof of $B$ from $A_1, \ldots, A_m$, there is an $\mathcal{M}_L$-proof of $[\![B]\!]$ from $[\![A_1]\!], \ldots, [\![A_m]\!]$.
- $\mathcal{M}_L$ is *faithful for* $L$ if $\mathcal{M}_L$ is sound and complete for $L$.

Informally, $\mathcal{M}_{\text{IPL}}$ is sound for IPL because the additional axioms are true and the rules of $\mathcal{M}$ are sound. A better argument is by induction on normal proofs in $\mathcal{M}$. Here is a summary of the proof-theoretic concepts of Prawitz [31, 32]. For simplicity, let us ignore equality rules, identifying terms that are equivalent up to $\lambda$-conversions.

A *branch* in a proof traces the construction and destruction of a formula. Each branch is obtained by repeatedly walking downwards from a premise of a rule to its conclusion, but terminates at the second premise of $\Rightarrow$-elimination. Thus in

$$\frac{\phi \Rightarrow \psi \qquad \phi}{\psi}$$

a branch may connect $\phi \Rightarrow \psi$ with $\psi$ but not $\phi$ with $\psi$ since these formulae may be syntactically unrelated. (This discussion is for $\mathcal{M}$. For logics having other connectives, most elimination rules are special cases.)

Every proof in $\mathcal{M}$ can be *normalized* such that, in every branch, no elimination rule immediately follows an introduction rule. In a normal proof, every branch

| | *introduction* (I) | *elimination* (E) |
|---|---|---|
| *Conjunction* | $\dfrac{A \qquad B}{A \,\&\, B}$ | $\dfrac{A \,\&\, B}{A}$ $\quad$ $\dfrac{A \,\&\, B}{B}$ |
| *Disjunction* | $\dfrac{A}{A \vee B}$ $\quad$ $\dfrac{B}{A \vee B}$ | $\dfrac{A \vee B \qquad \begin{matrix}[A] \\ C\end{matrix} \qquad \begin{matrix}[B] \\ C\end{matrix}}{C}$ |
| *Implication* | $\dfrac{\begin{matrix}[A] \\ B\end{matrix}}{A \supset B}$ | $\dfrac{A \supset B \qquad A}{B}$ |
| *Contradiction* | | $\dfrac{\bot}{A}$ |

Figure 1: The rules of intuitionistic propositional logic

$$\bigwedge AB \,.\, [\![A]\!] \Rightarrow ([\![B]\!] \Rightarrow [\![A \,\&\, B]\!]) \tag{\&I}$$

$$\bigwedge AB \,.\, [\![A \,\&\, B]\!] \Rightarrow [\![A]\!] \qquad \bigwedge AB \,.\, [\![A \,\&\, B]\!] \Rightarrow [\![B]\!] \tag{\&E}$$

$$\bigwedge AB \,.\, [\![A]\!] \Rightarrow [\![A \vee B]\!] \qquad \bigwedge AB \,.\, [\![B]\!] \Rightarrow [\![A \vee B]\!] \tag{$\vee$I}$$

$$\bigwedge ABC \,.\, [\![A \vee B]\!] \Rightarrow ([\![A]\!] \Rightarrow [\![C]\!]) \Rightarrow ([\![B]\!] \Rightarrow [\![C]\!]) \Rightarrow [\![C]\!] \tag{$\vee$E}$$

$$\bigwedge AB \,.\, ([\![A]\!] \Rightarrow [\![B]\!]) \Rightarrow [\![A \supset B]\!] \tag{$\supset$ I}$$

$$\bigwedge AB \,.\, [\![A \supset B]\!] \Rightarrow [\![A]\!] \Rightarrow [\![B]\!] \tag{$\supset$ E}$$

$$\bigwedge A \,.\, [\![\bot]\!] \Rightarrow [\![A]\!] \tag{$\bot$E}$$

Figure 2: Meta-level axioms for intuitionistic propositional logic

$$
\dfrac{\dfrac{\dfrac{\dfrac{\bigwedge AB \,.\, \llbracket A\rrbracket \Rightarrow (\llbracket B\rrbracket \Rightarrow \llbracket A \mathbin{\&} B\rrbracket)}{\bigwedge B \,.\, \llbracket C\rrbracket \Rightarrow (\llbracket B\rrbracket \Rightarrow \llbracket C \mathbin{\&} B\rrbracket)}}{\llbracket C\rrbracket \Rightarrow (\llbracket D\rrbracket \Rightarrow \llbracket C \mathbin{\&} D\rrbracket)} \qquad \begin{array}{c}\vdots\\ \llbracket C\rrbracket\end{array}}{\llbracket D\rrbracket \Rightarrow \llbracket C \mathbin{\&} D\rrbracket} \qquad \begin{array}{c}\vdots\\ \llbracket D\rrbracket\end{array}}{\llbracket C \mathbin{\&} D\rrbracket}
$$

Figure 3: The meta-proof formalizing a &I inference

$$
\dfrac{\dfrac{\dfrac{\bigwedge AB \,.\, (\llbracket A\rrbracket \Rightarrow \llbracket B\rrbracket) \Rightarrow \llbracket A \supset B\rrbracket}{\bigwedge B \,.\, (\llbracket C\rrbracket \Rightarrow \llbracket B\rrbracket) \Rightarrow \llbracket C \supset B\rrbracket}}{(\llbracket C\rrbracket \Rightarrow \llbracket D\rrbracket) \Rightarrow \llbracket C \supset D\rrbracket} \qquad \dfrac{\begin{array}{c}[\,\llbracket C\rrbracket\,]\\ \vdots\\ \llbracket D\rrbracket\end{array}}{\llbracket C\rrbracket \Rightarrow \llbracket D\rrbracket}}{\llbracket C \supset D\rrbracket}
$$

Figure 4: The meta-proof formalizing an $\supset$I inference

begins with an assumption or axiom, then has a series of eliminations, then a series of introductions. During the eliminations the formulae shrink to a minimum; during the introductions they grow again.

Observe that $\llbracket B\rrbracket$ is an atomic $\mathcal{M}_{\text{IPL}}$-formula. A normal proof can be put into *expanded* normal form, where every minimum formula is atomic [32, page 254]. For example, if a minimum formula is $\phi \Rightarrow \psi$, then the following can be spliced into the proof, reducing the minimum formula to $\psi$:

$$
\dfrac{\dfrac{\phi \Rightarrow \psi \qquad [\phi]}{\psi}}{\phi \Rightarrow \psi}
$$

Completeness holds because to each object-level inference there corresponds a meta-proof involving an $\mathcal{M}_{\text{IPL}}$ axiom. Soundness holds because to each occurrence of an $\mathcal{M}_{\text{IPL}}$ axiom in a meta-proof there corresponds an object-level inference. Figures 3 and 4 illustrate the correspondence.

**Theorem 2** $\mathcal{M}_{\text{IPL}}$ *is sound for* IPL.
*Proof*: By induction on the size of the expanded normal proof in $\mathcal{M}_{\text{IPL}}$ of $\llbracket B\rrbracket$ from $\llbracket A_1\rrbracket, \ldots, \llbracket A_m\rrbracket$, construct an IPL proof of $B$ from $A_1, \ldots, A_m$.

Since $\llbracket B\rrbracket$ is atomic, the branch terminating with $\llbracket B\rrbracket$ cannot contain introduction rules, and thus cannot discharge assumptions. The branch must consist entirely of elimination rules. If it is just $\llbracket B\rrbracket$ then $B$ is an assumption, one of $A_1, \ldots, A_m$. Otherwise the branch contains elimination rules, so its first formula cannot be atomic. It must consist of an axiom followed by elimination rules. There is one case for each axiom.

For the &I axiom, $B$ is $C \,\&\, D$ for some formulae $C$ and $D$. The meta-proof must have the structure of Figure 3. It has two $\bigwedge$-eliminations involving $C$ and $D$, and two $\Rightarrow$-eliminations, involving proofs of $[\![C]\!]$ and $[\![D]\!]$ from $[\![A_1]\!], \ldots, [\![A_m]\!]$. By the induction hypothesis, construct IPL proofs of $C$ and $D$ from $A_1, \ldots, A_m$. Applying &I gives an IPL proof of $C \,\&\, D$.

For the $\supset$I axiom, $B$ is $C \supset D$. The meta-proof must have the structure of Figure 4. It contains a proof of $[\![C]\!] \Rightarrow [\![D]\!]$ from $[\![A_1]\!], \ldots, [\![A_m]\!]$. By expanded normal form this consists of a proof of $[\![D]\!]$ from $[\![A_1]\!], \ldots, [\![A_m]\!], [\![C]\!]$, followed by $\Rightarrow$-introduction, discharging the assumption $[\![C]\!]$. By the induction hypothesis, construct an IPL proof of $D$ from $A_1, \ldots, A_m, C$, and $\supset$I gives an IPL proof of $C \supset D$ from $A_1, \ldots, A_m$.

The cases for the other axioms are similar. $\square$

**Theorem 3** $\mathcal{M}_{\text{IPL}}$ *is complete for* IPL.
*Proof*: By induction on the size of the IPL proof of $B$ from $A_1, \ldots, A_m$, construct a proof of $[\![B]\!]$ from $[\![A_1]\!], \ldots, [\![A_m]\!]$ in $\mathcal{M}_{\text{IPL}}$.

Suppose the last inference of the IPL proof is $\supset$I, and the conclusion is $C \supset D$. Then the rule is applied to an IPL proof of $D$ from $A_1, \ldots, A_m, C$. By the induction hypothesis, construct an $\mathcal{M}_{\text{IPL}}$-proof of $[\![D]\!]$ from $[\![A_1]\!], \ldots, [\![A_m]\!], [\![C]\!]$. Now it is easy to construct a meta-proof like that in Figure 4.

The cases for the other axioms are similar. $\square$

**Remark.** The heavy use of meta-implication ($\Rightarrow$) may suggest that the meta-logic ought to involve sequents. Then the &I axiom would be

$$[\![A]\!], [\![B]\!] \vdash [\![A \,\&\, B]\!]$$

and the resolution rule (Section 4.2) would resemble the cut rule. However, this amounts to using the sequent symbol as a logical connective. In typical usage, an axiom is a formula — not a sequent [36]. Such a use of sequents would also complicate the use of hypothetical rules (illustrated in Section 4.4).

# 4 Backwards proof construction

The reduction of the goal $\phi$ to the subgoals $\phi_1, \ldots, \phi_m$ amounts to deriving the rule $\phi_1, \ldots, \phi_m/\phi$. The meta-logic represents this object-level rule as the implication $[\phi_1, \ldots, \phi_m] \Rightarrow \phi$. Such Horn clauses are combined by a derived meta-rule: resolution.

Let us begin by looking at backwards proof as a style of proof construction. Then we can formalize backwards proof in $\mathcal{M}$, obtaining the methods implemented in Isabelle.

$$\frac{A \mathbin{\&} B \supset C \supset A \mathbin{\&} C}{A \mathbin{\&} B \supset C \supset A \mathbin{\&} C} \;\xmapsto[1]{\supset\mathrm{I}}\; \frac{\begin{array}{c} A \mathbin{\&} B \\ \vdots \\ C \supset A \mathbin{\&} C \end{array}}{A \mathbin{\&} B \supset C \supset A \mathbin{\&} C} \;\xmapsto[2]{\supset\mathrm{I}}\; \frac{\begin{array}{c} A \mathbin{\&} B \\ C \\ \vdots \\ A \mathbin{\&} C \end{array}}{A \mathbin{\&} B \supset C \supset A \mathbin{\&} C}$$

$$\xmapsto[3]{\&\mathrm{I}}\; \frac{\begin{array}{c} A \mathbin{\&} B \\ C \\ \vdots \\ A \end{array} \qquad \begin{array}{c} A \mathbin{\&} B \\ C \\ \vdots \\ C \end{array}}{A \mathbin{\&} B \supset C \supset A \mathbin{\&} C} \;\xmapsto[4]{\mathrm{asm}}\; \frac{\begin{array}{c} A \mathbin{\&} B \\ C \\ \vdots \\ A \end{array}}{A \mathbin{\&} B \supset C \supset A \mathbin{\&} C} \;\xmapsto[5]{\&\mathrm{E}}\; A \mathbin{\&} B \supset C \supset A \mathbin{\&} C$$

Figure 5: The steps of the backwards proof

## 4.1 Proof states as derived rules

In this section, 'rule,' 'proof,' etc., refer to object-level rules and proofs. The method is illustrated by a sample proof in intuitionistic propositional logic:

$$\cfrac{\cfrac{\cfrac{\cfrac{[A \mathbin{\&} B]}{A} \qquad [C]}{A \mathbin{\&} C}}{C \supset A \mathbin{\&} C}}{A \mathbin{\&} B \supset (C \supset A \mathbin{\&} C)}$$

A backwards proof grows from the root upwards, rather than from the leaves downwards. Every state of the proof can be represented by a derived rule whose conclusion is the main goal, here $A \mathbin{\&} B \supset (C \supset A \mathbin{\&} C)$, and whose premises are the current subgoals. The proof of the derived rule — namely, the internal structure — has no further role and is suppressed. The initial state is represented by the trivial rule whose premise and conclusion are identical.

Figure 5 shows the sequence of proof states, represented as derived rules. Each step is written as an arrow (like $\xmapsto[2]{\supset\mathrm{I}}$) giving the inference rule and step number. The initial rule is combined with $\supset$I, to derive a rule involving the discharge of the assumption $A \mathbin{\&} B$. Combining this rule again with $\supset$I adds $C$ to the assumptions. Combining the resulting rule with &I derives a rule that has two premises. This splits the goal $A \,\&\, C$ in two. The full proof at this point is

$$\cfrac{\cfrac{\cfrac{\begin{array}{cc} A \,\&\, B & A \,\&\, B \\ C & C \\ \vdots & \vdots \\ A & C \end{array}}{A \,\&\, C}}{C \supset A \,\&\, C}}{A \,\&\, B \supset (C \supset A \,\&\, C)}$$

The second subgoal, $C$, holds trivially by assumption. The first subgoal is proved using &E to prove $A$ from $A\&B$. The final state is the theorem $A\&B \supset (C \supset A\&C)$.

## 4.2 Proof construction by resolution

To formalize this kind of proof construction in $\mathcal{M}$ requires a change in notation. An object-inference like $\frac{A\&B}{A}$ will now be written $[\![A \,\&\, B]\!] \Rightarrow [\![A]\!]$. The meta-inference $\phi \longmapsto \psi$ will now be written $\frac{\phi}{\psi}$.

The initial state in a proof of $C$ is formalized by the trivial meta-theorem $[\![C]\!] \Rightarrow [\![C]\!]$. A state of the proof having $n$ subgoals is represented by a derived object-rule having $n$ premises, formalized by the meta-theorem

$$[\psi_1, \ldots, \psi_n] \Rightarrow [\![C]\!]$$

The state with zero subgoals is the meta-theorem $[\![C]\!]$, which represents the object-theorem $C$. Outer quantifiers are dropped, so the meta-theorems may contain free variables.

Most Isabelle proof steps use resolution, a derived meta-rule. Resolution instantiates free variables of an object-rule by unification against a subgoal in the proof state. In a simple case, the substitution $s$ must match $\phi$ against $\psi$, namely $\phi s \equiv \psi$:

$$\frac{\Phi \Rightarrow \phi \qquad \Psi \Rightarrow \psi \Rightarrow \theta}{\Psi \Rightarrow \Phi s \Rightarrow \theta} \tag{1}$$

The list notation for nested implication eliminates subscripts; $\Phi$ and $\Psi$ are lists of formulae. Here is the meta-rule again, writing out the lists in full.

Let $i$ be given; if $\phi s \equiv \psi_i$ then

$$\frac{[\phi_1, \ldots, \phi_m] \Rightarrow \phi \qquad [\psi_1, \ldots, \psi_i, \ldots, \psi_n] \Rightarrow \theta}{[\psi_1, \ldots, \phi_1 s, \ldots, \phi_m s, \ldots, \psi_n] \Rightarrow \theta}$$

Resolution replaces the subgoal $\psi_i$ by $\phi_1 s, \ldots, \phi_m s$ in the proof state. The first premise is an object-level rule and the second is a proof state; the conclusion is a new proof state:

$$\frac{\text{object-level rule} \qquad \text{proof state}}{\text{new proof state}}$$

Full resolution involves unification. The free variables of the object-level rule should be distinct from those of the proof state. In the examples, variables are renamed by subscripting. Isabelle variables contain an index to facilitate renaming.

Resolution is easily derived in $\mathcal{M}$. Using both quantifier rules $k$ times derives an *instantiation* rule,

$$\frac{\phi}{\phi[a_1/x_1, \ldots, a_k/x_k]}$$

provided that $x_1, \ldots, x_k$ are not free in the assumptions. Resolution consists of instantiation followed by reasoning about implication.

**Remark.** Isabelle resolution [28] is ordinary resolution restricted to Horn clauses. Many Isabelle proof procedures use techniques of logic programming. Completeness of resolution is not a central issue in Isabelle, for a proof may use any meta-rules.

## 4.3 Lifting a rule over assumptions

The resolution rule (1) represents LCF-style backwards proof, which does not recognize natural deduction. In LCF (and early versions of Isabelle) natural deduction must be expressed through a sequent calculus. Our formalization of IPL expresses natural deduction by meta-implication. We now supplement resolution with a method (called *lifting*) of making assumptions at the meta-level.

Suppose we want to formalize the sample proof, Figure 5. Resolving the $\supset$I axiom against the goal $[\![A \,\&\, B \supset (C \supset A \,\&\, C)]\!]$ will produce the subgoal

$$[\![A \,\&\, B]\!] \Rightarrow [\![C \supset A \,\&\, C]\!] \tag{2}$$

which contains the assumption $A \,\&\, B$. The next backwards step should use $\supset$I again. We can prepare the axiom by replacing its bound variables $A$ and $B$ by new free variables $A_2$ and $B_2$,

$$([\![A_2]\!] \Rightarrow [\![B_2]\!]) \Rightarrow [\![A_2 \supset B_2]\!]$$

and then *lift* it over the assumption $A \,\&\, B$, to obtain the meta-theorem

$$([\![A \,\&\, B]\!] \Rightarrow ([\![A_2]\!] \Rightarrow [\![B_2]\!])) \Rightarrow ([\![A \,\&\, B]\!] \Rightarrow [\![A_2 \supset B_2]\!]) \tag{3}$$

Resolution with this instantiates $A_2$ and $B_2$, replacing the subgoal (2) by

$$[\![A \,\&\, B]\!] \Rightarrow ([\![C]\!] \Rightarrow [\![A \,\&\, C]\!])$$

This subgoal contains the assumptions $A \,\&\, B$ and $C$.

**The lifting rule.** The general case is lifting the object-rule $[\phi_1, \ldots, \phi_m] \Rightarrow \phi$ over the list of assumptions $\Theta$:

$$\frac{[\phi_1, \ldots, \phi_m] \Rightarrow \phi}{[\Theta \Rightarrow \phi_1, \ldots, \Theta \Rightarrow \phi_m] \Rightarrow (\Theta \Rightarrow \phi)}$$

This meta-rule, called lifting or $\Rightarrow$-lifting, is derivable in $\mathcal{M}$.

Suppose first that $\Theta = [\theta]$. Clearly $\Rightarrow$-introduction gives

$$\frac{\psi}{\theta \Rightarrow \psi}$$

If $\psi$ represents an object-rule then it must be an implication, say $\phi \Rightarrow \psi'$. To push $\theta$ to the right, assume $\theta \Rightarrow \phi$ and apply the following meta-rule, whose derivation is simple:

$$\frac{\theta \Rightarrow (\phi \Rightarrow \psi') \qquad \theta \Rightarrow \phi}{\theta \Rightarrow \psi'}$$

If $\psi'$ is also an implication then repeating this step pushes $\theta$ fully to the right; finally, an equal number of $\Rightarrow$-introductions discharges the assumptions like $\theta \Rightarrow \phi$. If $\psi'$ is not an implication then the result would be

$$(\theta \Rightarrow \phi) \Rightarrow (\theta \Rightarrow \psi')$$

Now if $\Theta = [\theta_1, \ldots, \theta_k]$ then repeatedly apply the process above to $\theta_k$, ..., $\theta_1$.

**A sample proof.** Now we can formalize the proof of $A \,\&\, B \supset (C \supset A \,\&\, C)$. The first step is the resolution of the $\supset$I axiom with the initial proof state:

$$\frac{([\![A_1]\!] \Rightarrow [\![B_1]\!]) \Rightarrow [\![A_1 \supset B_1]\!] \qquad [\![A \,\&\, B \supset C \supset A \,\&\, C]\!] \Rightarrow [\![A \,\&\, B \supset C \supset A \,\&\, C]\!]}{([\![A \,\&\, B]\!] \Rightarrow [\![C \supset A \,\&\, C]\!]) \Rightarrow [\![A \,\&\, B \supset C \supset A \,\&\, C]\!]}$$

The $\supset$I axiom was prepared for resolution by dropping outer quantifiers, introducing new free variables $A_1$ and $B_1$. Resolution instantiated $A_1$ to $A \,\&\, B$ and $B_1$ to $C \supset A \,\&\, C$. The new state has one subgoal: to prove $C \supset A \,\&\, C$ from the assumption $A \,\&\, B$.

Recall that lifting the $\supset$I axiom over $A \,\&\, B$ produces the meta-theorem (3). Resolving (3) with the proof state instantiates $A_2$ to $C$ and $B_2$ to $A \,\&\, C$:

$$\frac{(3) \qquad ([\![A \,\&\, B]\!] \Rightarrow [\![C \supset A \,\&\, C]\!]) \Rightarrow [\![A \,\&\, B \supset (C \supset A \,\&\, C)]\!]}{([\![A \,\&\, B]\!] \Rightarrow [\![C]\!] \Rightarrow [\![A \,\&\, C]\!]) \Rightarrow [\![A \,\&\, B \supset (C \supset A \,\&\, C)]\!]}$$

To save space, define abbreviations for the main goal and the assumption list:

$$\begin{aligned} \psi &= [\![A \,\&\, B \supset (C \supset A \,\&\, C)]\!] \\ \Theta &= [\,[\![A \,\&\, B]\!], [\![C]\!]\,] \end{aligned}$$

Thus the current proof state is $(\Theta \Rightarrow [\![A \,\&\, C]\!]) \Rightarrow \psi$.

The next step is resolution with the &I axiom, after lifting it over the assumptions:

$$\frac{(\Theta \Rightarrow [\![A_3]\!]) \Rightarrow (\Theta \Rightarrow [\![B_3]\!]) \Rightarrow (\Theta \Rightarrow [\![A_3 \,\&\, B_3]\!]) \qquad (\Theta \Rightarrow [\![A \,\&\, C]\!]) \Rightarrow \psi}{(\Theta \Rightarrow [\![A]\!]) \Rightarrow (\Theta \Rightarrow [\![C]\!]) \Rightarrow \psi}$$

The variable instantiations are $A_3$ to $A$ and $B_3$ to $C$. Observe how the assumptions, $\Theta$, are copied to both subgoals.

The next step solves the second goal by assumption. It is formalized by resolution with the meta-tautology $\Theta \Rightarrow [\![C]\!]$, *without* lifting:

$$\frac{\Theta \Rightarrow [\![C]\!] \qquad (\Theta \Rightarrow [\![A]\!]) \Rightarrow (\Theta \Rightarrow [\![C]\!]) \Rightarrow \psi}{(\Theta \Rightarrow [\![A]\!]) \Rightarrow \psi}$$

Next, resolution with the &E axiom reduces the goal $A$ to the subgoal $A \,\&\, B$. Since the goal cannot instantiate the variable $B$ of the &E axiom, let us take the correct instance of the axiom, lifted over $\Theta$:

$$\frac{(\Theta \Rightarrow [\![A \,\&\, B]\!]) \Rightarrow (\Theta \Rightarrow [\![A]\!]) \qquad (\Theta \Rightarrow [\![A]\!]) \Rightarrow \psi}{(\Theta \Rightarrow [\![A \,\&\, B]\!]) \Rightarrow \psi}$$

Full resolution (with unification) can instantiate variables in the proof state. But Section 4.4 gives a better treatment of &E.

We solve the last subgoal by assumption — resolution with another meta-tautology:

$$\frac{\Theta \Rightarrow [\![A \,\&\, B]\!] \qquad (\Theta \Rightarrow [\![A \,\&\, B]\!]) \Rightarrow \psi}{\psi}$$

This concludes the proof of $A \,\&\, B \supset (C \supset A \,\&\, C)$, representing each step of the object-level proof by a meta-level resolution. The proof is valid for arbitrary object-formulae $A$, $B$, and $C$. To emphasize this we may generalize $\psi$ over its free variables, obtaining the final meta-theorem

$$\bigwedge ABC \,.\, [\![A \,\&\, B \supset (C \supset A \,\&\, C)]\!]$$

## 4.4 Deriving object-level rules

Gordon [13] prefers higher-order logic because it can express many of its derived rules as theorems. Isabelle expresses derived rules as meta-theorems, allowing us to work in first-order logic or even weaker systems.

**A new conjunction rule.** The &E rules typically work in the forwards direction: from $A \,\&\, B$ conclude $A$. Another version of conjunction elimination, resembling $\vee$E, is better suited to backwards proof:

$$\frac{A \,\&\, B \qquad \begin{matrix}[A, B] \\ C\end{matrix}}{C}$$

This conjunction rule discharges the assumptions $A$ and $B$ in its second premise; we can also say [33] that its second premise is the rule $\frac{A \quad B}{C}$.

The formalization in $\mathcal{M}_{\mathrm{IPL}}$ is

$$\bigwedge ABC \,.\, [\![A \,\&\, B]\!] \Rightarrow ([\![A]\!] \Rightarrow [\![B]\!] \Rightarrow [\![C]\!]) \Rightarrow [\![C]\!]$$

To prove it, start with the goal $[\![C]\!]$. For fixed $A$, $B$, $C$ we may assume $[\![A \,\&\, B]\!]$ and $[\![A]\!] \Rightarrow [\![B]\!] \Rightarrow [\![C]\!]$. Once $[\![C]\!]$ has been proved, discharging the meta-assumptions and generalizing produces the correct meta-theorem.

The first step is resolution with the meta-assumption $[\![A]\!] \Rightarrow [\![B]\!] \Rightarrow [\![C]\!]$:

$$\frac{[\![A]\!] \Rightarrow [\![B]\!] \Rightarrow [\![C]\!] \qquad [\![C]\!] \Rightarrow [\![C]\!]}{[\![A]\!] \Rightarrow [\![B]\!] \Rightarrow [\![C]\!]}$$

The subgoals are $[\![A]\!]$ and $[\![B]\!]$. We resolve &E against the first subgoal:

$$\frac{[\![A \,\&\, B]\!] \Rightarrow [\![A]\!] \qquad [\![A]\!] \Rightarrow [\![B]\!] \Rightarrow [\![C]\!]}{[\![A \,\&\, B]\!] \Rightarrow [\![B]\!] \Rightarrow [\![C]\!]}$$

Next, resolve with the meta-assumption $[\![A \,\&\, B]\!]$:

$$\frac{[\![A \,\&\, B]\!] \qquad [\![A \,\&\, B]\!] \Rightarrow [\![B]\!] \Rightarrow [\![C]\!]}{[\![B]\!] \Rightarrow [\![C]\!]}$$

Next, use the other &E axiom:

$$\frac{[\![A \,\&\, B]\!] \Rightarrow [\![B]\!] \qquad [\![B]\!] \Rightarrow [\![C]\!]}{[\![A \,\&\, B]\!] \Rightarrow [\![C]\!]}$$

Finally, resolve with $[\![A \,\&\, B]\!]$ again:

$$\frac{[\![A \,\&\, B]\!] \qquad [\![A \,\&\, B]\!] \Rightarrow [\![C]\!]}{[\![C]\!]}$$

**Hypothetical rules.** Adding the double negation rule to intuitionistic propositional logic gives classical logic. More precisely, if every formula $A$ satisfies the double negation rule,

$$\frac{(A \supset \bot) \supset \bot}{A}$$

then every formula $B$ satisfies the excluded middle, $B \vee (B \supset \bot)$.[2]

The formalization of this entailment is

$$(\bigwedge A \,.\, [\![(A \supset \bot) \supset \bot]\!] \Rightarrow [\![A]\!]) \Rightarrow (\bigwedge B \,.\, [\![B \vee (B \supset \bot)]\!])$$

Here the premise is a schematic rule. We must make, and finally discharge, the quantified meta-assumption

$$\bigwedge A \,.\, [\![(A \supset \bot) \supset \bot]\!] \Rightarrow [\![A]\!]$$

Conversely, any formula that satisfies the excluded middle also satisfies double negation [11, page 27]:

$$\frac{A \vee (A \supset \bot) \qquad (A \supset \bot) \supset \bot}{A}$$

The formalization of this derived rule is

$$\bigwedge A \,.\, [\![A \vee (A \supset \bot)]\!] \Rightarrow [\![(A \supset \bot) \supset \bot]\!] \Rightarrow [\![A]\!]$$

Contrast the use of quantifiers in these examples.

[2]The proof is tricky. From $(B \vee (B \supset \bot)) \supset \bot$ prove both $B \supset \bot$ and $(B \supset \bot) \supset \bot$.

# 5 Quantification

Many logical constants introduce bound variables: universal and existential quantifiers ($\forall$ and $\exists$), description operators ($\lambda$, $\iota$ and $\epsilon$), general product and sum ($\Pi$ and $\Sigma$), union and intersection of families (as in $\bigcup_{i \in I} A_i$), and so on. Isabelle implements logics comprising most of these.

Adding quantifiers to the previous object-logic gives intuitionistic first-order logic (IFOL). Formally, extend $\mathcal{M}_{\text{IPL}}$ to become $\mathcal{M}_{\text{IFOL}}$. Add the type *term* for denotations of terms. The quantifiers are the constant symbols

$$\forall, \exists \quad : \quad (\mathit{term} \to \mathit{form}) \to \mathit{form}$$

If $A$, $A(x)$, and $A(x, y)$ each have type *form* then the three variables named $A$ must have different types, and so are different variables. Rather than declaring a fixed list of variables with their types, let the context determine the types — avoiding things like $A \,\&\, A(x)$. For emphasis, $F$, $G$, and $H$ will stand for formula-valued functions.

Write $\forall x.A$ for $\forall(\lambda x.A)$ and $\exists x.A$ for $\exists(\lambda x.A)$. By $\lambda$-conversion every quantified formula is equivalent to one of the form $\forall(F)$ or $\exists(F)$, where $F$ has type $\mathit{term} \to \mathit{form}$.

The rules (Figure 6) and their meta-level axioms (Figure 7) do not have the close resemblance that we saw for propositional logic. The eigenvariable conditions of $\forall$I and $\exists$E are not formalized literally. Note that the two conditions differ in form but not in effect. Both ensure that $x$ serves only to specify a truth-valued function, through its occurrences in $A$.

In the axioms, $F$ denotes not the text of the quantification but its meaning: a truth-valued function. The axiom $\forall$I states that if $F$ is an everywhere-true function then $\forall x.F(x)$ is true. Similarly, $B$ denotes not the text of a formula but a truth-value. The $\exists$E axiom states that if $\exists x.F(x)$ is true and $F(x)$ implies $B$ for all $x$, then $B$ is true. The axioms reflect the meanings of the corresponding rules.

Although the justification of each axiom is semantic, they behave as expected in syntax. Substitution for the variables $F$ and $B$ avoids capture of the variable $x$. In particular, $B$ may not be replaced by a formula containing $x$. Assumptions also obey the eigenvariable conditions, as we shall see below.

The demonstration that these axioms faithfully represent first-order logic is similar to that for propositional logic (Section 3).

**Theorem 4** $\mathcal{M}_{\text{IFOL}}$ *is sound for* IFOL.
*Proof*: By induction over the expanded normal proof in $\mathcal{M}_{\text{IFOL}}$ of $[\![B]\!]$ from $[\![A_1]\!], \ldots, [\![A_m]\!]$, construct an IFOL proof of $B$ from $A_1, \ldots, A_m$. The branch terminating with $[\![B]\!]$, unless it is trivial, consists of an axiom followed by elimination rules.

For the $\exists$I axiom, $B$ is $\exists x.G(x)$. The normalized proof must have the form shown in Figure 8. Two $\bigwedge$-eliminations introduce $G$ and $u$; then $\Rightarrow$-elimination is applied to a proof of $[\![G(u)]\!]$ from $[\![A_1]\!], \ldots, [\![A_m]\!]$. By the induction hypothesis construct an IFOL proof of $G(u)$ from $A_1, \ldots, A_m$, and use the $\exists$I rule to prove $\exists x.G(x)$. The

| | *introduction* (I) | *elimination* (E) |
|---|---|---|
| *Universal quantifier* | $\dfrac{A}{\forall x.A}\,*$ | $\dfrac{\forall x.A}{A[t/x]}$ |
| *Existential quantifier* | $\dfrac{A[t/x]}{\exists x.A}$ | $\dfrac{\exists x.A \qquad \begin{matrix}[A]\\ B\end{matrix}}{B}\,*$ |

**Eigenvariable conditions*:
$\forall$I: provided $x$ not free in the assumptions
$\exists$E: provided $x$ not free in $B$ or in any assumption save $A$

Figure 6: Quantifier rules

$$\bigwedge F\,.\,(\bigwedge x\,.\,\llbracket F(x)\rrbracket) \Rightarrow \llbracket \forall x.F(x)\rrbracket \tag{$\forall$I}$$

$$\bigwedge F y\,.\,\llbracket \forall x.F(x)\rrbracket \Rightarrow \llbracket F(y)\rrbracket \tag{$\forall$E}$$

$$\bigwedge F y\,.\,\llbracket F(y)\rrbracket \Rightarrow \llbracket \exists x.F(x)\rrbracket \tag{$\exists$I}$$

$$\bigwedge F B\,.\,\llbracket \exists x.F(x)\rrbracket \Rightarrow (\bigwedge x\,.\,\llbracket F(x)\rrbracket \Rightarrow \llbracket B\rrbracket) \Rightarrow \llbracket B\rrbracket \tag{$\exists$E}$$

Figure 7: Meta-level axioms for the quantifier rules

$$\dfrac{\dfrac{\dfrac{\bigwedge F y\,.\,\llbracket F(y)\rrbracket \Rightarrow \llbracket \exists x.F(x)\rrbracket}{\bigwedge y\,.\,\llbracket G(y)\rrbracket \Rightarrow \llbracket \exists x.G(x)\rrbracket}}{\llbracket G(u)\rrbracket \Rightarrow \llbracket \exists x.G(x)\rrbracket} \qquad \begin{matrix}\vdots\\ \llbracket G(u)\rrbracket\end{matrix}}{\llbracket \exists x.G(x)\rrbracket}$$

Figure 8: The meta-proof formalizing an $\exists$I inference

$$\dfrac{\dfrac{\dfrac{\dfrac{\bigwedge F B\,.\,\llbracket \exists x.F(x)\rrbracket \Rightarrow (\bigwedge x\,.\,\llbracket F(x)\rrbracket \Rightarrow \llbracket B\rrbracket) \Rightarrow \llbracket B\rrbracket}{\bigwedge B\,.\,\llbracket \exists x.G(x)\rrbracket \Rightarrow (\bigwedge x\,.\,\llbracket G(x)\rrbracket \Rightarrow \llbracket B\rrbracket) \Rightarrow \llbracket B\rrbracket}}{\llbracket \exists x.G(x)\rrbracket \Rightarrow (\bigwedge x\,.\,\llbracket G(x)\rrbracket \Rightarrow \llbracket C\rrbracket) \Rightarrow \llbracket C\rrbracket} \qquad \begin{matrix}\vdots\\ \llbracket \exists x.G(x)\rrbracket\end{matrix}}{(\bigwedge x\,.\,\llbracket G(x)\rrbracket \Rightarrow \llbracket C\rrbracket) \Rightarrow \llbracket C\rrbracket} \qquad \dfrac{\dfrac{\begin{matrix}[\,\llbracket G(y)\rrbracket\,]\\ \vdots\\ \llbracket C\rrbracket\end{matrix}}{\llbracket G(y)\rrbracket \Rightarrow \llbracket C\rrbracket}}{\bigwedge y\,.\,\llbracket G(y)\rrbracket \Rightarrow \llbracket C\rrbracket}}{\llbracket C\rrbracket}$$

Figure 9: The meta-proof formalizing an $\exists$E inference

$\mathcal{M}_{\text{IFOL}}$ proof is shown without $\beta$-conversions, identifying terms that have the same normal form. If $G$ is $\lambda x.A$ then $G(u) \equiv A[u/x]$, and $[\![\exists x.G(x)]\!] \equiv [\![\exists x.A]\!]$.

For $\exists$E, the proof (Figure 9) contains a proof of $\bigwedge y \,.\, [\![G(y)]\!] \Rightarrow [\![C]\!]$ from $[\![A_1]\!]$, $\ldots$, $[\![A_m]\!]$. Assuming expanded normal form, it consists of a proof of $[\![C]\!]$ followed by $\Rightarrow$-introduction, discharging $[\![G(y)]\!]$, followed by $\bigwedge$-introduction. (The bound variable $y$ can be chosen so that it is not free in $[\![A_1]\!], \ldots, [\![A_m]\!]$.) By the induction hypothesis, there are IFOL proofs of $C$ from $A_1, \ldots, A_m, G(y)$ and of $\exists x.G(x)$ from $A_1, \ldots, A_m$. The $\exists$E rule gives an IFOL proof of $C$ from $A_1, \ldots, A_m$.

The cases for the other axioms are similar. □

**Theorem 5** *$\mathcal{M}_{\text{IFOL}}$ is complete for* IFOL.
*Proof*: By induction over the IFOL proof of $B$ from $A_1, \ldots, A_m$, construct a proof of $[\![B]\!]$ from $[\![A_1]\!], \ldots, [\![A_m]\!]$ in $\mathcal{M}_{\text{IFOL}}$.

The hardest case is when the last inference is $\exists$E. Then the rule is applied to an IFOL proof of $\exists x.A$, and to a proof of $B$ from $A$. By the axiom for $\exists$E, it is enough to prove the theorems $[\![\exists x.A]\!]$ and $\bigwedge x \,.\, [\![A]\!] \Rightarrow [\![B]\!]$. By the induction hypothesis, there is an $\mathcal{M}_{\text{IFOL}}$-proof of $[\![\exists x.A]\!]$, and also a proof of $[\![B]\!]$ from $[\![A]\!]$. The meta-proof resembles that in Figure 9, where $G$ is $\lambda x.A$. Again, terms having the same normal form are identified. □

**Remark.** Perhaps the type names *term* and *form* are overly syntactic; *term* denotes a set of individuals while *form* denotes a set of truth-values. The meaning of $A \supset B$ should depend on the meanings of $A$ and $B$, not on their syntactic structure.

Still, types play an important syntactic role. An expression of type *term* represents an IFOL term, and similarly *form* represents formulae. By assigning a type to each syntactic category of the object-logic, type-checking in $\mathcal{M}$ enforces syntactic constraints.

# 6 Backwards proof with quantifiers

Quantifiers complicate backwards proof: goals may contain unknowns and parameters. An *unknown* takes the form of a free variable in the proof state, and can be replaced by any term. Unification is a standard technique for solving unknowns. A *parameter* comes from an eigenvariable of a rule. Parameters have two possible representations, $\bigwedge$-bound variables and Hilbert $\epsilon$-terms.

## 6.1 Lifting over universal quantifiers

The universal quantifier $\bigwedge$ is the obvious way to express a goal involving parameters. Recall the axioms $\forall$I and $\exists$E. If we use $\forall$I to prove $\forall z.G(z) \vee H(z)$ then the subgoal will be $\bigwedge z \,.\, [\![G(z) \vee H(z)]\!]$. Resolution against a quantified goal requires a new derived meta-rule: lifting an object-rule over a quantified variable. This $\bigwedge$-lifting resembles $\Rightarrow$-lifting, introduced earlier to handle assumptions.

Suppose we want to apply $\vee$I to the goal $\bigwedge z\,.\,[\![G(z) \vee H(z)]\!]$. The $\vee$I axiom has quantified variables $A$ and $B$:

$$\bigwedge AB\,.\,[\![A]\!] \Rightarrow [\![A \vee B]\!]$$

(Lifting seems easier to explain if quantifiers are shown explicitly. In the sample proofs below, outer quantifiers are dropped after lifting.) Lifting the axiom over $z$ produces a meta-theorem with new quantified variables $G$ and $H$, functions of $z$:

$$\bigwedge GH\,.\,(\bigwedge z\,.\,[\![G(z)]\!]) \Rightarrow (\bigwedge z\,.\,[\![G(z) \vee H(z)]\!]) \qquad (4)$$

Why is this a theorem? By the $\vee$I axiom, $[\![G(z)]\!]$ implies $[\![G(z) \vee H(z)]\!]$ for all $G$, $H$, and $z$. Thus we may reduce $\bigwedge z\,.\,[\![G(z) \vee H(z)]\!]$ to the subgoal $\bigwedge z\,.\,[\![G(z)]\!]$.

The $\exists$I axiom has quantified variables $F$ and $y$:

$$\bigwedge Fy\,.\,[\![F(y)]\!] \Rightarrow [\![\exists x.F(x)]\!]$$

Lifting over $z$ produces quantified variables $G$ and $f$:

$$\bigwedge Gf\,.\,(\bigwedge z\,.\,[\![G(z, f(z))]\!]) \Rightarrow (\bigwedge z\,.\,[\![\exists x.G(z, x)]\!]) \qquad (5)$$

By the $\exists$I axiom, $[\![G(z, f(z))]\!]$ implies $[\![\exists x.G(z, x)]\!]$ for all $G$, $f$, and $z$. Thus we may reduce $\bigwedge z\,.\,[\![\exists x.G(z, x)]\!]$ to the subgoal $\bigwedge z\,.\,[\![G(z, f(z))]\!]$. The pattern may become clearer if we recall that $\exists x.F(x)$ means $\exists(F)$ and $\exists x.G(z, x)$ means $\exists(G(z))$; we have replaced $F$ by $G(z)$.

**The lifting rule.** Given an object-rule, lifting replaces all the outer quantified variables by new ones of function type. Its formal derivation consists of several steps. Consider the $\mathcal{M}$-proof

$$\cfrac{\cfrac{\bigwedge y\,.\,\psi}{\psi[f(z)/y]}}{\bigwedge z\,.\,\psi[f(z)/y]}$$

Here $f$ is a *variable* (of function type), and the final step will involve generalization over $f$. For $k$ quantified variables we can similarly derive the rule

$$\frac{\bigwedge y_1 \ldots y_k\,.\,\psi}{\bigwedge z\,.\,\psi[f_1(z)/y_1, \ldots, f_k(z)/y_k]}$$

where the $f_i$ are free variables. If $z$ has type $\sigma$ then lifting replaces $y_i$ (of type $\tau_i$) by a different variable $f_i$ (of type $\sigma \to \tau_i$).

Typically $\psi$ is an implication $\phi \Rightarrow \psi'$. To push the $\bigwedge z$ into the implication, assume $\bigwedge z\,.\,\phi$ and apply this derived meta-rule:

$$\frac{\bigwedge z\,.\,\phi \Rightarrow \psi' \qquad \bigwedge z\,.\,\phi}{\bigwedge z\,.\,\psi'}$$

If the object-rule has several premises then $\psi'$ will also be an implication; continue pushing the $\bigwedge z$ into the right. Then use $\Rightarrow$-introduction to discharge the assumptions like $\bigwedge z\,.\,\phi$, and use $\bigwedge$-introduction to generalize over the variables $f_1$, …, $f_k$. This yields general forms of lifting.

Repeated lifting over the variables $z_n$, …, $z_1$ allows resolution with the goal $\bigwedge z_1 \ldots z_n.\theta$.

**A sample proof.** Consider a proof of $\forall z.G(z) \vee H(z)$ from $\forall z.G(z)$:

$$\cfrac{\cfrac{\cfrac{\forall z.G(z)}{G(z)}}{G(z) \vee H(z)}}{\forall z.G(z) \vee H(z)}$$

In the meta-proof, $G$ and $H$ are function variables. We may afterwards substitute $\forall z.A$ for $G$ and $\forall z.B$ for $H$, where $A$ and $B$ are any formulae.

Working backwards, the first inference is $\forall$I. The first resolution instantiates $F_1$ to $\lambda z.G(z) \vee H(z)$:

$$\frac{(\bigwedge z\,.\,[\![F_1(z)]\!]) \Rightarrow [\![\forall z.F_1(z)]\!] \qquad\qquad [\![\forall z.G(z) \vee H(z)]\!] \Rightarrow [\![\forall z.G(z) \vee H(z)]\!]}{(\bigwedge z\,.\,[\![G(z) \vee H(z)]\!]) \Rightarrow [\![\forall z.G(z) \vee H(z)]\!]}$$

Here resolution must cope with function variables like $F_1$, determining the correct $\lambda$-abstraction and identifying terms that have the same normal form. For this Isabelle uses higher-order unification [19, 28]. The conclusion is normalized to eliminate $\beta$-redexes.

Observe that the goal $G(z) \vee H(z)$ must be proved for arbitrary $z$, a parameter. The next step uses the $\vee$I axiom, which becomes the meta-theorem (4) after lifting over $z$. Resolution with this produces the subgoal $G(z)$:

$$\frac{(\bigwedge z.[\![G_2(z)]\!]) \Rightarrow \bigwedge z.[\![G_2(z) \vee H_2(z)]\!] \qquad (\bigwedge z.[\![G(z) \vee H(z)]\!]) \Rightarrow [\![\forall z.G(z) \vee H(z)]\!]}{(\bigwedge z\,.\,[\![G(z)]\!]) \Rightarrow [\![\forall z.G(z) \vee H(z)]\!]}$$

The instantiations are $G_2$ to $G$ and $H_2$ to $H$.

The next inference, $\forall$E, cannot easily be used backwards. Its conclusion, $A[t/x]$, can match a goal in numerous ways. Higher-order unification of a term with $F(y)$, where $F$ and $y$ are variables, yields many unifiers.[3] Instead, work forwards. Making the assumption $[\![\forall z.G(z)]\!]$, the $\forall$E axiom proves $\bigwedge y.[\![G(y)]\!]$. (Recall that we are deriving a rule, and therefore may assume its premise.) Lifting this theorem over $z$ gives $\bigwedge f.\bigwedge z\,.\,[\![G(f(z))]\!]$.

$$\frac{\bigwedge z\,.\,[\![G(f_3(z))]\!] \qquad (\bigwedge z\,.\,[\![G(z)]\!]) \Rightarrow [\![\forall z.G(z) \vee H(z)]\!]}{[\![\forall z.G(z) \vee H(z)]\!]}$$

The final resolution replaces $f_3$ by $\lambda z.z$, and so

$$G(f_3(z)) \equiv G((\lambda z.z)(z)) \equiv G(z)$$

This use of function variables is typical: the term $f_3(z)$ states precisely its dependence on the context. The following sections give more examples.

[3] When this example was run, Isabelle did actually make the correct unifier its first choice.

**Both kinds of lifting together.** The goal $\forall z\,.\,P(z) \supset P(z) \vee Q(z)$ leads to the proof state

$$(\bigwedge z\,.\,[\![P(z)]\!] \Rightarrow [\![P(z) \vee Q(z)]\!]) \Rightarrow [\![\forall z\,.\,P(z) \supset P(z) \vee Q(z)]\!]$$

The next step requires both kinds of lifting: to lift the $\vee$I axiom over the assumption $P(z)$ and then the variable $z$. The result after lifting is

$$\bigwedge GH\,.\,(\bigwedge z\,.\,P(z) \Rightarrow [\![G(z)]\!]) \Rightarrow (\bigwedge z\,.\,P(z) \Rightarrow [\![G(z) \vee H(z)]\!])$$

The methods of this section and 4.3 work together to derive this sort of lifting.

## 6.2 Unification

Unification is a powerful aid for reasoning about quantification. We can prove $a < b$ by proving $a < c$ and $c < b$ for some $c$. A procedure runs in linear time provided that, for some constant $K$, if the input has size $n$ then the run time is $Kn$. Given $\forall x.P(x) \vee Q(x)$ we may argue by cases on whether $P(a)$ or $Q(a)$ holds, leaving $a$ unspecified. Each of these cases involve unknowns: terms that must eventually be stated to complete the proof. Unification instantiates unknowns in goals.

Certain examples in first-order logic are especially helpful. Construct a proof of $\exists xy.P(f(x), y)$ from $\forall z.P(z, g(z))$, considering how unification can determine the instantiations of $x$, $y$, and $z$. To exercise all the quantifier rules, prove $\forall x.\exists y.P(x, y)$ from $\exists y.\forall x.P(x, y)$ and try to prove the converse. Here we will work two simpler examples: a proof of $\forall x.\exists y\,.\,x = y$, and an attempted proof of $\exists y.\forall x\,.\,x = y$. These illustrate eigenvariable conditions.

The examples involve a reflexive equality relation, with the axiom $\bigwedge y\,.\,[\![y = y]\!]$.

**Resolution using unification.** Resolution easily handles goals containing unknowns. Simply extend the meta-rule (1) of Section 4.2 to instantiate *both* premises, the object-rule and the proof state. If the substitution $s$ satisfies $\phi s \equiv \psi s$ then

$$\frac{\Phi \Rightarrow \phi \qquad \Psi \Rightarrow \psi \Rightarrow \theta}{\Psi s \Rightarrow \Phi s \Rightarrow \theta s}$$

This resolution rule applies if $s$ unifies $\phi$ and $\psi$. Resolution can affect variables throughout the proof state: $\Psi$ becomes $\Psi s$ and $\theta$ becomes $\theta s$.

**A successful proof.** A first-order proof of $\forall x.\exists y\,.\,x = y$ is

$$\cfrac{\cfrac{x = x}{\exists y\,.\,x = y}}{\forall x.\exists y\,.\,x = y}$$

The first resolution in the backwards proof of $\forall x.\exists y\,.\,x = y$ involves the $\forall$I axiom, instantiating $F_1$ to $\lambda x.\exists y\,.\,x = y$:

$$\frac{(\bigwedge x\,.\,[\![F_1(x)]\!]) \Rightarrow [\![\forall x.F_1(x)]\!] \qquad [\![\forall x.\exists y\,.\,x = y]\!] \Rightarrow [\![\forall x.\exists y\,.\,x = y]\!]}{(\bigwedge x\,.\,[\![\exists y\,.\,x = y]\!]) \Rightarrow [\![\forall x.\exists y\,.\,x = y]\!]}$$

The meta-theorem (5) is the result of lifting the ∃I axiom over a variable. We use the lifted axiom here with different variable names. Resolution yields

$$\frac{(\bigwedge x\,.\,[\![G_2(x,f_2(x))]\!]) \Rightarrow \bigwedge x\,.\,[\![\exists y.G_2(x,y)]\!] \qquad (\bigwedge x\,.\,[\![\exists y.x=y]\!]) \Rightarrow [\![\forall x.\exists y.x=y]\!]}{(\bigwedge x\,.\,[\![x=f_2(x)]\!]) \Rightarrow [\![\forall x.\exists y.x=y]\!]}$$

The variable $G_2$ is instantiated to $\lambda xy\,.\,x=y$, so the normal form of $G_2(x,f_2(x))$ is $x=f_2(x)$.

Putting $\lambda x.x$ for $f_2$ solves the subgoal, $\bigwedge x\,.\,[\![x=f_2(x)]\!]$, by reflexivity. In the formal proof, this happens by lifting the reflexivity axiom over $x$ and then resolving:

$$\frac{\bigwedge x\,.\,[\![g_3(x)=g_3(x)]\!] \qquad (\bigwedge x\,.\,[\![x=f_2(x)]\!]) \Rightarrow [\![\forall x.\exists y\,.\,x=y]\!]}{[\![\forall x.\exists y\,.\,x=y]\!]}$$

Consider the steps of higher-order unification [19, 28]. The initial disagreement pair is

$$\langle \bigwedge x\,.\,[\![g_3(x)=g_3(x)]\!],\ \bigwedge x\,.\,[\![x=f_2(x)]\!]\rangle$$

It reduces to the pairs $\langle \lambda x.g_3(x), \lambda x.x\rangle$ and $\langle \lambda x.g_3(x), \lambda x.f_2(x)\rangle$. The first pair forces $g_3(x)$ to be $\lambda x.x$; the second forces $f_2(x)$ to be $\lambda x.x$. This is a unifier: the common instance is $\bigwedge x\,.\,[\![x=x]\!]$.

**An unsuccessful proof.** An attempt to prove $\exists y.\forall x\,.\,x=y$ is

$$\cfrac{\cfrac{x=t}{\forall x\,.\,x=t}}{\exists y.\forall x\,.\,x=y}$$

Here $t$ can be any term not containing $x$ free. No such term satisfies $x=t$, so the top formula is false.

The first resolution in the attempted proof of $\exists y.\forall x\,.\,x=y$ involves the ∃I axiom:

$$\frac{[\![F_1(y_1)]\!] \Rightarrow [\![\exists x.F_1(x)]\!] \qquad [\![\exists y.\forall x\,.\,x=y]\!] \Rightarrow [\![\exists y.\forall x\,.\,x=y]\!]}{[\![\forall x\,.\,x=y_1]\!] \Rightarrow [\![\exists y.\forall x\,.\,x=y]\!]}$$

The subgoal contains a new variable, $y_1$.

Resolution with the ∀I axiom gives

$$\frac{(\bigwedge x\,.\,[\![F_2(x)]\!]) \Rightarrow [\![\forall x.F_2(x)]\!] \qquad [\![\forall x\,.\,x=y_1]\!] \Rightarrow [\![\exists y.\forall x\,.\,x=y]\!]}{(\bigwedge x\,.\,[\![x=y_1]\!]) \Rightarrow [\![\exists y.\forall x\,.\,x=y]\!]}$$

We are stuck. The subgoal $\bigwedge x\,.\,[\![x=y_1]\!]$ is false; no fixed $y_1$ can equal every $x$. Resolution with the reflexivity axiom fails. The initial disagreement pair

$$\langle \bigwedge x\,.\,[\![g_3(x)=g_3(x)]\!],\ \bigwedge x\,.\,[\![x=y_1]\!]\rangle$$

reduces to the pairs $\langle \lambda x.g_3(x), \lambda x.x\rangle$ and $\langle \lambda x.g_3(x), \lambda x.y_1\rangle$. The first pair forces $g_3(x)$ to be $\lambda x.x$, reducing the second to $\langle \lambda x.x, \lambda x.y_1\rangle$, which has no unifier.

# 7 Other representations of eigenvariables

We have seen how to express most forms of Isabelle-86 proof construction as formal inference in a meta-logic. What about the Isabelle-86 treatment of eigenvariables?

Isabelle-86 does not use $\bigwedge$-lifting; it enforces eigenvariable conditions literally. The quantifier rule

$$\frac{\Gamma \models A(y)}{\Gamma \models \forall x.A(x)}$$

holds provided $y$ is not free in $\Gamma$ or $A$. (Isabelle-86 cannot handle natural deduction, so its object-logics are typically sequent calculi.) Isabelle-86 reduces $\Gamma \models \forall x.A(x)$ to the subgoal $\Gamma \models A(y)$, where $y$ is a variable not previously used in the proof. Since two proofs must not be combined if they have eigenvariables in common, resolution renames all eigenvariables in its first premise. Isabelle-86 maintains the eigenvariable conditions — it accounts for assignments to variables in $\Gamma$ and $A$, forbidding those that would introduce $y$ into $\Gamma$ or $A$.

The meta-logic $\mathcal{M}$ represents each inference rule as an implication. This fails if a premise has free variables. Subject to the usual conditions, the following inference is valid:

$$\frac{\phi[y/x]}{\bigwedge x.\phi}$$

The following implication is *invalid*:

$$\phi[y/x] \Rightarrow \bigwedge x.\phi \tag{6}$$

Both cases involve implicit quantification over $y$, but with different scopes.

Replacing $y$ by a special term can transform the implication (6) into a valid formula. For this, $\mathcal{M}$ is too weak a meta-logic, so let us temporarily adopt full higher-order logic.

## 7.1 Hilbert's $\epsilon$-operator

Church's formulation of higher-order logic includes the Hilbert $\epsilon$-operator: $\epsilon x.\psi$ is a term for any formula $\psi$. It embodies a strong Axiom of Choice. If $\psi$ is true for some value of $x$ then $\epsilon x.\psi$ denotes some such value; otherwise $\epsilon x.\psi$ has an arbitrary value of the same type as $x$. (Recall that no type may denote the empty set.) An axiom scheme for the $\epsilon$-operator is

$$\bigwedge x \,.\, \psi \Rightarrow \psi[(\epsilon x.\psi)\,/\,x]$$

Full higher-order logic has classical negation. Putting $\neg\phi$ for $\psi$, taking the contrapositive, and pushing the $\bigwedge x$ inwards, gives

$$\phi[(\epsilon x.\neg\phi)\,/\,x] \Rightarrow \bigwedge x.\phi$$

So the special term to plug into the implication (6) is $\epsilon x.\neg\phi$: a value chosen to falsify $\phi$, if such exists.

The term $\epsilon x.\neg\phi$ contains the same free variables as $\bigwedge x.\phi$. For example, a theorem representing ∀I in first-order logic is

$$\bigwedge F\,.\,[\![\,F(\epsilon x\,.\,\neg[\![F(x)]\!])\,]\!] \Rightarrow [\![\forall x.F(x)]\!]$$

where the variable $F$ is free in $\epsilon x\,.\,\neg[\![F(x)]\!]$ but bound in the surrounding term. Specializing $F$ to $\lambda x\,.\,g(x) = 0$ yields the theorem

$$[\![g(\epsilon x\,.\,\neg[\![g(x) = 0]\!]) = 0]\!] \Rightarrow [\![\forall x\,.\,g(x) = 0]\!]$$

Specializing $F$ to $\lambda x\,.\,G(x) \supset H(x)$ yields the theorem

$$[\![G(\epsilon x\,.\,\neg[\![G(x) \supset H(x)]\!]) \supset H(\epsilon x\,.\,\neg[\![G(x) \supset H(x)]\!])]\!] \Rightarrow [\![\forall x\,.\,G(x) \supset H(x)]\!]$$

The term $\epsilon x\,.\,\neg[\![F(x)]\!]$ produces two different terms

$$\epsilon x\,.\,\neg[\![g(x) = 0]\!] \qquad \text{and} \qquad \epsilon x\,.\,\neg[\![G(x) \supset H(x)]\!]$$

Each eigenvariable is represented by a giant term: $\epsilon x\,.\,\neg[\![F(x)]\!]$ contains the formula $F$, which may contain other $\epsilon$-terms. I tried this cumbersome representation in early versions of Isabelle, but no method of structure sharing would control the exponential growth.

## 7.2 Replacing Hilbert's $\epsilon$ by special constants

Here is something similar to $\epsilon$-terms but more practical. Extend higher-order logic with the axiom scheme

$$\phi[\mathbf{y}\{\phi\}\,/\,x] \Rightarrow \bigwedge x.\phi \qquad (7)$$

We simply postulate constants to satisfy the implication (6). For each $\phi$ there is a *unique* constant $\mathbf{y}\{\phi\}$ not free in $\phi$. This is Henkin's technique for reducing first-order logic to propositional logic [5, page 30]. Its main application is the Completeness Theorem; here it extends our propositional proof methods to first-order logic.

The simplest representation of $\mathbf{y}\{\phi\}$ is to regard $\phi$ as part of the name, but this is as cumbersome as $\epsilon$-terms. Enumerating names like $\mathbf{y}\{609\}$ (Lisp's GENSYM) does not work if $\bigwedge x.\phi$ contains free variables: substitution could create distinct instances of (7), and even introduce occurrences of $\mathbf{y}\{609\}$ in $\bigwedge x.\phi$. We must make certain that one constant is not used for different instances of (7) in a proof.

Taking the free variables of $\bigwedge x.\phi$ as part of the name, and updating them, provides an automatic renaming mechanism — precisely that of Isabelle-86. Thus $\mathbf{y}\{\phi\}$ should be written $\mathbf{y}\{z_1, \ldots, z_n\}$, where $z_1,\ldots,z_n$ are the free variables of $\bigwedge x.\phi$. The only way to introduce occurrences of $\mathbf{y}\{\phi\}$ into $\bigwedge x.\phi$ is to instantiate some free variable $z_i$ with $\mathbf{y}\{z_1, \ldots, z_n\}$. This is prevented by the occurs check in unification.

The $\epsilon$-terms can express functions like $\lambda x\,.\,\epsilon y\,.\,\psi$, which maps $x$ to some $y$ such that $\psi$ holds. But $\mathbf{y}\{z_1, \ldots, z_n\}$ is a *constant* — so $z_1$, …, $z_n$ must not be bound in the surrounding term. This restricts the generalization rule ($\bigwedge$-introduction) to variables not mentioned by special constants. Instead we can use the instantiation rule,

$$\frac{\phi}{\phi[a_1/x_1, \ldots, a_k/x_k]}$$

where the variables $x_1$, ..., $x_k$ must not be free in the assumptions. In $\mathcal{M}$ the rule follows by generalization and specialization, but here it must be taken as primitive. In each special constant, $x_i$ is replaced by all free variables of $a_i$.

The ∀I axiom becomes

$$[\![F(\mathbf{y}\{F\})]\!] \Rightarrow [\![\forall x.F(x)]\!]$$

Note that there is no $\bigwedge F$: since the variable $F$ is mentioned by the constant $\mathbf{y}\{F\}$, it must be free in the axiom. However, we may replace $F$ by a term, creating a new special constant. Replacing $F$ by $\lambda x \,.\, g(x) = 0$ yields the theorem

$$[\![g(\mathbf{y}\{g\}) = 0]\!] \Rightarrow [\![\forall x \,.\, g(x) = 0]\!]$$

Replacing $F$ by $\lambda x.G(x) \supset H(x)$ yields the theorem

$$[\![G(\mathbf{y}\{G,H\}) \supset H(\mathbf{y}\{G,H\})]\!] \Rightarrow [\![\forall x.G(x) \supset H(x)]\!]$$

The theorems contain different constants $\mathbf{y}\{g\}$ and $\mathbf{y}\{G,H\}$: thus $\mathbf{y}\{F\}$ is automatically renamed.

Compare these with the $\epsilon$-term examples. Keeping just the free variables of $F$, rather than $F$ itself, prevents the exponential growth.

## 7.3 Lifting versus special constants

Redoing the examples from Section 6.2 illustrates how special constants work.

**A successful proof.** The first resolution in the backwards proof of $\forall x.\exists y \,.\, x = y$ involves the ∀I axiom:

$$\frac{[\![F_1(\mathbf{x}\{F_1\})]\!] \Rightarrow [\![\forall x.F_1(x)]\!] \qquad [\![\forall x.\exists y \,.\, x = y]\!] \Rightarrow [\![\forall x.\exists y \,.\, x = y]\!]}{[\![\exists y \,.\, \mathbf{x}\{\} = y]\!] \Rightarrow [\![\forall x.\exists y \,.\, x = y]\!]}$$

Here $F_1$ is assigned $\lambda x.\exists y \,.\, x = y$, which has no free variables. Thus the special constant $\mathbf{x}\{\}$ in the conclusion has no variables.

Next we resolve with the ∃I axiom:

$$\frac{[\![F_2(y_2)]\!] \Rightarrow [\![\exists x.F_2(x)]\!] \qquad [\![\exists y \,.\, \mathbf{x}\{\} = y]\!] \Rightarrow [\![\forall x.\exists y \,.\, x = y]\!]}{[\![\mathbf{x}\{\} = y_2]\!] \Rightarrow [\![\forall x.\exists y \,.\, x = y]\!]}$$

The function variable $F_2$ is instantiated to $\lambda y \,.\, \mathbf{x}\{\} = y$, so the normal form of $F_2(y_2)$ is $\mathbf{x}\{\} = y_2$.

Resolving the proof state with the reflexivity axiom sets $y_2$ to $\mathbf{x}\{\}$, completing the proof:

$$\frac{[\![x_3 = x_3]\!] \qquad [\![\mathbf{x}\{\} = y_2]\!] \Rightarrow [\![\forall x.\exists y \,.\, x = y]\!]}{[\![\forall x.\exists y \,.\, x = y]\!]}$$

**A unsuccessful proof.** The first step in the attempted proof of $\exists y.\forall x\,.\,x = y$ is the same as before:

$$\frac{[\![F_1(y_1)]\!] \Rightarrow [\![\exists x.F_1(x)]\!] \qquad [\![\exists y.\forall x\,.\,x = y]\!] \Rightarrow [\![\exists y.\forall x\,.\,x = y]\!]}{[\![\forall x\,.\,x = y_1]\!] \Rightarrow [\![\exists y.\forall x\,.\,x = y]\!]}$$

Resolution with the $\forall$I axiom gives

$$\frac{[\![F_2(\mathbf{x}\{F_2\})]\!] \Rightarrow [\![\forall x.F_2(x)]\!] \qquad [\![\forall x\,.\,x = y_1]\!] \Rightarrow [\![\exists y.\forall x\,.\,x = y]\!]}{[\![\mathbf{x}\{y_1\} = y_1]\!] \Rightarrow [\![\exists y.\forall x\,.\,x = y]\!]}$$

The instantiation of $F_2$ is $\lambda x\,.\,x = y_1$, which contains free variable $y_1$: hence the special constant $\mathbf{x}\{y_1\}$ in the conclusion. The subgoal $[\![\mathbf{x}\{y_1\} = y_1]\!]$ cannot be solved because the instantiation of $y_1$ to $\mathbf{x}\{y_1\}$ would be circular. These terms are not unifiable.

**Parallel derivations.** A final example will compare $\bigwedge$-lifting with special constants. The main goal, $\psi$, is

$$[\![\exists u.\forall x.\exists v.\forall y.\exists w.P(u, x, v, y, w)]\!]$$

Each state in the backwards proof has one subgoal. Here are the derivations for both styles, in parallel:

| *lifting* | *special constants* |
|---|---|
| $[\![\forall x.\exists v.\forall y.\exists w.P(a, x, v, y, w)]\!] \Rightarrow \psi$ | $[\![\forall x.\exists v.\forall y.\exists w.P(a, x, v, y, w)]\!] \Rightarrow \psi$ |
| $(\bigwedge x.[\![\exists v.\forall y.\exists w.P(a, x, v, y, w)]\!]) \Rightarrow \psi$ | $[\![\exists v.\forall y.\exists w.P(a, \mathbf{x}\{a\}, v, y, w)]\!] \Rightarrow \psi$ |
| $(\bigwedge x.[\![\forall y.\exists w.P(a, x, f(x), y, w)]\!]) \Rightarrow \psi$ | $[\![\forall y.\exists w.P(a, \mathbf{x}\{a\}, b, y, w)]\!] \Rightarrow \psi$ |
| $(\bigwedge xy.[\![\exists w.P(a, x, f(x), y, w)]\!]) \Rightarrow \psi$ | $[\![\exists w.P(a, \mathbf{x}\{a\}, b, \mathbf{y}\{a, b\}, w)]\!] \Rightarrow \psi$ |
| $(\bigwedge xy.[\![P(a, x, f(x), y, g(x, y))]\!]) \Rightarrow \psi$ | $[\![P(a, \mathbf{x}\{a\}, b, \mathbf{y}\{a, b\}, c)]\!] \Rightarrow \psi$ |

In the sequence of states new symbols appear one by one: the free variable $a$, the parameter $x$, the free variable $f$ or $b$, the parameter $y$, and the free variable $g$ or $c$. Consider the final meta-theorem (the last line above):

| *lifting* | *special constants* |
|---|---|
| The variable $a$ cannot be assigned a term containing $x$ or $y$ free because $x$ and $y$ are bound variables. | The variable $a$ cannot be assigned a term containing $\mathbf{x}\{a\}$ or $\mathbf{y}\{a, b\}$ because $a$ is free in these constants. |
| Assigning $\lambda x.b$ to $f$ replaces the term $f(x)$ by $b$, which may contain $x$ free but not $y$. | The variable $b$ cannot be assigned a term containing $\mathbf{y}\{a, b\}$, for this constant contains $b$. |
| Assigning $\lambda xy.c$ to $g$ replaces the term $g(x, y)$ by $c$, which may contain $x$ and $y$ free. | The variable $c$ can be assigned a term containing $\mathbf{x}\{a\}$ or $\mathbf{y}\{a, b\}$, for $c$ is free in neither constant. |

Special constants raise many questions. When is the name of a special constant significant, and what is its scope? Does the axiom scheme (7) entail classical logic or the Axiom of Choice? With lifting, the status of parameters is clear. They can be renamed by $\alpha$-conversion. The scope of a parameter is its goal. Here is a proof state with two subgoals:

$$(\bigwedge x.\phi) \Rightarrow (\bigwedge y.\theta) \Rightarrow \psi$$

The scope of $x$ is the first goal and that of $y$ is the second.

Lifting is clearly preferable — especially because it works in the simple meta-logic $\mathcal{M}$. Higher-order unification easily copes with the additional function variables like $f$ and $g$ above.

# 8 An implementation

This research has always been concerned with implementation issues, of which the main one is how to represent a logic. I have criticized LCF for representing inference rules as functions, preferring visible structures [28]. But colleagues have pointed out that such structures are essentially the theorems of a (meta) logic, and the functions that manipulate them are (meta) rules. Thus Isabelle adopts LCF's representation at the meta-level.

Object-rules in Isabelle-86 are expressed in the typed $\lambda$-calculus and have the familiar pattern of premises over conclusion. Free variables and special constants give the effect of weak quantification. Written in $\mathcal{M}$, the general form of an Isabelle-86 rule is something like

$$(\bigwedge y_1.\phi_1) \Rightarrow \cdots \Rightarrow (\bigwedge y_m.\phi_m) \Rightarrow \phi$$

Below the level of rules, Isabelle operates on terms: substitution, normalization, higher-order unification, parsing and printing. Above, Isabelle is concerned with backwards proof: tactics and tacticals. Above pure Isabelle come the object-logics with their special tactics.

M. J. C. Gordon uses higher-order logic as a specification language for hardware. His HOL theorem prover [13] has supported several large proofs about circuit designs [6]. Since HOL is based on LCF, it may seem to be the obvious starting point for a version of Isabelle based on $\mathcal{M}$. However HOL and Isabelle support distinct methods of use, giving conflicting design requirements. I obtained the current version of Isabelle by modifying Isabelle-86. Only the level of rules needed substantial change; the lower levels were slightly modified and the higher levels hardly at all.

The natural deduction rules of $\mathcal{M}$ are represented by the corresponding sequent calculus. A meta-theorem has the form $\Psi \models \phi$, where $\phi$ depends on the assumptions $\Psi$, and expresses object-rules using $\bigwedge$ and $\Rightarrow$. Lifting and resolution are coded directly; implementing them by execution of primitive rules would be painfully slow.

Isabelle now uses $\bigwedge$-lifting instead of special constants. Despite having quantifiers, Isabelle provides a separate class of *schematic variables*[4] — free variables

[4]'Logical variables' in PROLOG jargon.

intended for substitution. Why have two kinds of free variables? Consider trying to prove

$$[\![(\exists z.F(z))\ \&\ B]\!] \Rightarrow [\![\exists z.F(z)\ \&\ B]\!]$$

which expresses a derived rule about $\exists$ and &. In the proof, the free variables $F$ and $B$ are fixed while the rightmost existential quantifier produces a schematic variable that must eventually be replaced by a term. The meta-rules ensure that every theorem $\Psi \models \phi$ has no schematic variables in $\Psi$; instantiation only affects schematic variables.

Object-logics include intuitionistic first-order logic, Martin-Löf's Constructive Type Theory (CTT) [23] and a first-order sequent calculus similar to LK [36]. The Isabelle-86 logics were easily adapted to the new Isabelle. The most drastic change in the representation of proofs, $\bigwedge$-lifting, was the main source of problems. But many things worked immediately, including the object-logics' specialized proof procedures. The Isabelle-86 logics were sequent calculi, but most now have a simpler formulation using natural deduction. Sequent calculi are mainly suitable for classical logic [32, page 245].

All proofs in this paper are in single steps. But Isabelle provides tacticals (resembling LCF's) for joining simple proof procedures into complex ones. Constructive Type Theory has a rewriting tactic that works by executing a 'logic program' derived from CTT rules. This tactic is heavily used in the largest CTT example, which develops elementary number theory up to the theorem

$$a \bmod b + (a/b) \times b = a$$

The first-order sequent calculus has an automatic proof procedure that performs associative unification using higher-order unification [20, page 37]. The procedure is not complete but it saves many steps in interactive proof and works with any convenient mixture of primitive and derived rules. Using derived rules about set theory, Isabelle can prove theorems like

$$c \neq \emptyset \models \bigcap_{x \in c}(f(x) \cap g(x)) = (\bigcap_{x \in c} f(x)) \cap (\bigcap_{x \in c} g(x))$$

Huet's unification procedure [19] may return, with a unifier, unsolved disagreement pairs $\langle a_1, b_1\rangle$, ..., $\langle a_m, b_m\rangle$. When this happens, the pairs have at least one and perhaps infinitely many unifiers. To avoid enumerating these unifiers the disagreement pairs can be kept to constrain future unifications. (This is Huet's key observation, which makes his procedure practical.) Isabelle-86 stores with each derived rule $\phi$ its unsolved disagreement pairs. Isabelle now expresses these using equality:

$$a_1 \equiv b_1 \Rightarrow \cdots \Rightarrow a_m \equiv b_m \Rightarrow \phi$$

Equality also replaces the Isabelle-86 definition mechanism. The meta-axiom $K \equiv a$ defines the constant $K$. We can then unfold $K$ to $a$, or fold $a$ back to $K$. This is more sensible than automatically unfolding all constants.

See the User's Manual for more detailed information about Isabelle [30].

# 9 Related work

De Bruijn's AUTOMATH is an early attempt (begun in 1966) to develop a formal language of mathematics [7]. The AUTOMATH languages are meant to appear natural to mathematicians while allowing the computer checking of proofs. Coquand and Huet's Calculus of Constructions [9] is similar but has a different treatment of abbreviations and context. Both are extended typed $\lambda$-calculi providing the Cartesian product of a family of types.[5]

Many pieces of mathematics have been computer checked in AUTOMATH [21] and the Constructions [10]. The mathematician specifies all items of discourse — including the logical constants — by writing axioms and definitions. The calculi can express classical reasoning, constructive reasoning, and various shades in between. But the properties of the logical constants may be unclear, for their definitions are technical.

Is there a clear, uniform formalization of logic? Martin-Löf's *theory of expressions* represents the syntax of his Constructive Type Theory. It is essentially Church's typed $\lambda$-calculus, which represents the syntax of higher-order logic. So Isabelle has always represented object-theorems in the typed $\lambda$-calculus. But what about the representation of rules? Initially I adopted an obvious form of inference rule, without fully developing the concept. As it happened, Schroeder-Heister had already done so.

With Schroeder-Heister's *rules of higher levels*, a rule may be the premise of another rule. This extends natural deduction: assumptions may be (higher level) rules, not just formulae [33]. Schematic variables in a rule express quantification [34]. Unfortunately, propositional logic requires a complicated treatment of assumption discharge, while the full conception is extremely complex. The meta-logic $\mathcal{M}$ is perhaps a rendering of Schroeder-Heister's work into a more convenient and familiar notation.

Martin-Löf has formulated a general concept of rule [24]. In a notation resembling AUTOMATH's he has produced succinct descriptions of first-order logic and Constructive Type Theory [25]. Most recently, people at the University of Edinburgh have elaborated Martin-Löf's ideas into the Edinburgh Logical Framework [15], and formalized diverse logics [3].

AUTOMATH and its descendants exploit the interpretation of *propositions-as-types*. Each proposition $A$ is interpreted as the type (or set) $\widetilde{A}$ of *proof objects*, and $A$ is true if it has a proof object, namely if there exists $a : \widetilde{A}$. Here is a summary of propositions-as-types [18, 27]:

- Absurdity ($\perp$) is interpreted as the empty type.
- A proof object of $A \,\&\, B$ consists of a pair $\langle a, b\rangle$, where $a$ is a proof object of $A$ and $b$ is a proof object of $B$ — namely where $a : \widetilde{A}$ and $b : \widetilde{B}$. Thus $A \,\&\, B$ is interpreted as the Cartesian product $\widetilde{A} \times \widetilde{B}$.

[5]The so-called 'dependent product' or 'dependent function space'

- A proof object of $A \vee B$ contains either a proof object of $A$ or else a proof object of $B$, and an indication of which. Thus $A \vee B$ is interpreted as the disjoint union $\widetilde{A} + \widetilde{B}$.

- A proof object of $A \supset B$ consists of a function $f$ such that if $a$ is a proof object of $A$ then $f(a)$ is a proof object of $B$ — namely if $a : \widetilde{A}$ then $f(a) : \widetilde{B}$. Thus $A \supset B$ is interpreted as the function type $\widetilde{A} \to \widetilde{B}$.

- A proof object of $\forall x{:}A \, . \, B(x)$ consists of a function $f$ such that if $a$ belongs to $A$ then $f(a)$ is a proof object of $B(a)$, namely if $a : A$ then $f(a) : \widetilde{B}(a)$. Thus $\forall x{:}A \, . \, B(x)$ is interpreted as the type $\Pi x{:}A \, . \, \widetilde{B}(x)$, the product of the family of types $\{\widetilde{B}(x)\}_{x:A}$.

Observe that if the type $B$ does not involve $x$ then $\{B\}_{x:A}$ is the constant family, and the product $\Pi x{:}A \, . \, B$ is the function type $A \to B$. But the $A$ in $A \supset B$ is a set $\widetilde{A}$ of proof objects, while the $A$ in $\forall x{:}A \, . \, B(x)$ typically is an ordinary set, like the natural numbers. Calculi like AUTOMATH are expressive and yet compact, for $\Pi$ encompasses function types, implication, and universal quantification.

The meta-logic $\mathcal{M}$, based on higher-order logic, is strikingly different. Its function types $\sigma \to \tau$, implications $\phi \Rightarrow \psi$, and universal quantifications $\bigwedge x.\phi$ are independent. Recall how each is used to formalize propositional logic. The type of the constant & indicates that it is a function on truth values. The axioms about & express properties of that function.

$$\& : \mathit{form} \to (\mathit{form} \to \mathit{form}) \qquad\qquad \bigwedge AB \, . \, [\![A]\!] \Rightarrow ([\![B]\!] \Rightarrow [\![A \,\&\, B]\!])$$

Propositions-as-types seems to offer advantages. Observe that $\bigwedge$-lifting and $\Rightarrow$-lifting would become simply $\Pi$-lifting. The richer type system allows a neat formalization of typed object-logics: the type $\mathit{term}(T)$ might represent the set of terms of object-type $T$. The formalization of a typed logic in $\mathcal{M}$ involves a type of all object-terms, including those of no legal object-type; object-level type checking requires additional rules for type inference. My choice of an old-fashioned calculus for Isabelle reflects practical considerations.

Under propositions-as-types, the formal proof of the proposition $A$ consists of a proof of $a : \widetilde{A}$, where $a$ is a proof object. A proof object is smaller than a standard proof tree (which contains repeated subformulae) yet still can grow large. Following the LCF tradition, Isabelle has never stored the steps of proofs. Milner's representation of logic makes stored proofs unnecessary, a vital space savings. Now some people regard this as a mistake. Martin-Löf's Type Theory exploits propositions-as-types; its proof objects constitute a functional programming language; the theorem prover *Nuprl* can execute them [8]. The proof objects of ordinary logics may be useful in proof editors.

Isabelle is intended for verifications involving hundreds of proofs, each involving hundreds of steps. Isabelle must support a degree of automation, and this requires unification. Propositions-as-types could consume excessive space; and what would take the place of Huet's unification procedure for higher-order logic [19]?

$$\frac{\dfrac{\forall x.A}{A}}{\exists x.A} \qquad \frac{\dfrac{\forall x.A}{A[0/x]}}{\exists x.A}$$

$(a)$ $\qquad$ $(b)$

Figure 10: Two proofs of $\exists x.A$ from $\forall x.A$

The choice of meta-logic can have subtle consequences stemming from the semantics. In $\mathcal{M}$ every type denotes a non-empty set; under propositions-as-types there must be an empty type. Standard first-order logic assumes a non-empty universe, and so $\forall x.A$ implies $\exists x.A$. The meta-logic $\mathcal{M}$ can represent the standard proof, Figure 10($a$). Avron et al. [3] suggest that deriving this in their formalization of first-order logic requires a constant, say 0. Their proof presumably represents that of Figure 10($b$).

A logic for the formalization of mathematics must presuppose the very minimum. Philosophers have debated whether a logic must be *predicative* — free of 'vicious circles' [37, page 37]:

> The vicious circles in question arise from supposing that a collection of objects may contain members which can only be defined by means of the collection as a whole. Thus, for example, the collection of *propositions* will be supposed to contain a proposition stating that "all propositions are either true or false." It would seem, however, that such a statement could not be legitimate unless "all propositions" referred to some already definite collection, which it cannot do if new propositions are created by statements about "all propositions." We shall, therefore, have to say that statements about "all propositions" are meaningless.

Whitehead and Russell's ramified type theory is the ancestor of Church's simple type theory (higher-order logic), which is impredicative. The simplest impredicative formula is $\bigwedge \theta . \theta$, a definition of absurdity: all propositions are true. The disjunction $\phi \vee \psi$ is equivalent to

$$\bigwedge \theta \,.\, (\phi \Rightarrow \theta) \Rightarrow (\psi \Rightarrow \theta) \Rightarrow \theta$$

Leibniz's definition of equality, $x = y$ if $y$ has every property of $x$, involves quantification over propositional functions: $\bigwedge \phi \,.\, \phi(x) \Rightarrow \phi(y)$.

Most of the other calculi are predicative and intuitionistic. To make $\mathcal{M}$ predicative, we can forbid bound variables whose type involves *prop*. The resulting system is much weaker but can still serve as a meta-logic, which requires only quantification over *object*-formulae. Its proof theory is as simple as that of first-order logic.

Amy Felty and Dale Miller [12] have formalized several logics in a logic programming language based on HOL. Logically their formalizations resemble those in Sections 3 and 5, but also express search procedures involving tactics and tacticals. Felty and Miller compare these with the usual approach of programming tactics in ML. They claim that the Edinburgh Logical Framework can be easily translated into higher-order logic.

Higher-order logic makes an adequate meta-logic from the theoretical perspective. We can draw on established proof theory to demonstrate soundness and completeness of the formalization of first-order logic: compare with the argument in Harper et al. [15]. The implementation, Isabelle, demonstrates its practical strengths. Basing Isabelle's reasoning methods on a precise calculus has already lead to extensions such as hypothetical rules. It is shedding light on issues such as theory structure.

**Acknowledgements.** Thierry Coquand contributed greatly to this research, which owes much to Michael Gordon's work on higher-order logic. David Matthews has continued to support his ML compiler. Philippe de Groote wrote the Isabelle-86 version of LK. Brian Monahan, David Wolfram and the referee commented on the paper. Thanks also to Peter Dybjer, Furio Honsell, Martin Hyland, Philippe Noël, and the editor. The SERC provided funding under grant GR/E0355.7.